%% file: paper_new.tex
\documentclass[sigplan,10pt,letter]{acmart}
\usepackage{amsmath,amssymb,amsfonts,amsthm}
\usepackage{algorithmic}
\usepackage{url}
\usepackage{paralist}
\usepackage{microtype}
\usepackage{paralist}
\usepackage{graphicx}
\usepackage{textcomp}
\usepackage{tabularx}
\usepackage{booktabs}
\usepackage{xcolor}
\usepackage{subfig}
\usepackage{listings}
\usepackage{tikz}
\usepackage{pgfplots}
\pgfplotsset{width=7cm,compat=1.8}

\usetikzlibrary{arrows,shapes,automata,fit}
\input{tumci}

\setcopyright{rightsretained}

\renewcommand\footnotetextcopyrightpermission[1]{} 
\acmYear{2019}
\copyrightyear{2019}

\definecolor{shadecolor}{gray}{.95}                                                           

\begin{document}

\title{Measuring Software Performance on Linux}
\subtitle{Technical Report\\\small\today}


\author{Martin Becker}
\orcid{0000-0003-3195-0503}
\affiliation{%
  \institution{Chair of Real-Time Computer Systems\\Technical University of Munich}
  \streetaddress{Arcisstrasse 21}
  \city{Munich}
  \state{Germany}
  \postcode{80333}
}
\email{martin.becker@tum.de}

\author{Samarjit Chakraborty}
\affiliation{%
  \institution{Chair of Real-Time Computer Systems\\Technical University of Munich}
  \streetaddress{Arcisstrasse 21}
  \city{Munich}
  \state{Germany}
  \postcode{80333}
}
\email{martin.becker@tum.de}



\begin{abstract}
  Measuring and analyzing the performance of software has reached a
  high complexity, caused by more advanced processor designs and the
  intricate interaction between user programs, the operating system,
  and the processor's microarchitecture.  In this report, we summarize
  our experience on how performance characteristics of software
  should be measured when running on a Linux operating system and a
  modern processor. In particular,
  \begin{inparaenum}[(1)]
  \item we provide a general overview about hardware and operating
    system features that may have a significant impact on timing and
    explain their interaction,
  \item we identify sources of errors that need to be controlled in
    order to obtain unbiased measurement results, and
  \item we propose a measurement setup for Linux to minimize errors.
  \end{inparaenum}
  Although not the focus of this report, we describe the measurement
  process using hardware performance counters, which can faithfully
  point to performance bottlenecks on a given processor.  Our experiments
  confirm that our measurement setup has a large impact on the
  results.  More surprisingly, however, they also suggest that the
  setup can be negligible for certain analysis methods. Furthermore,
  we found that our setup maintains significantly better performance
  under background load conditions, which means it can be used to
  improve high-performance applications.
\end{abstract}




  \begin{teaserfigure}
    \centering
    \includegraphics[height=8cm]{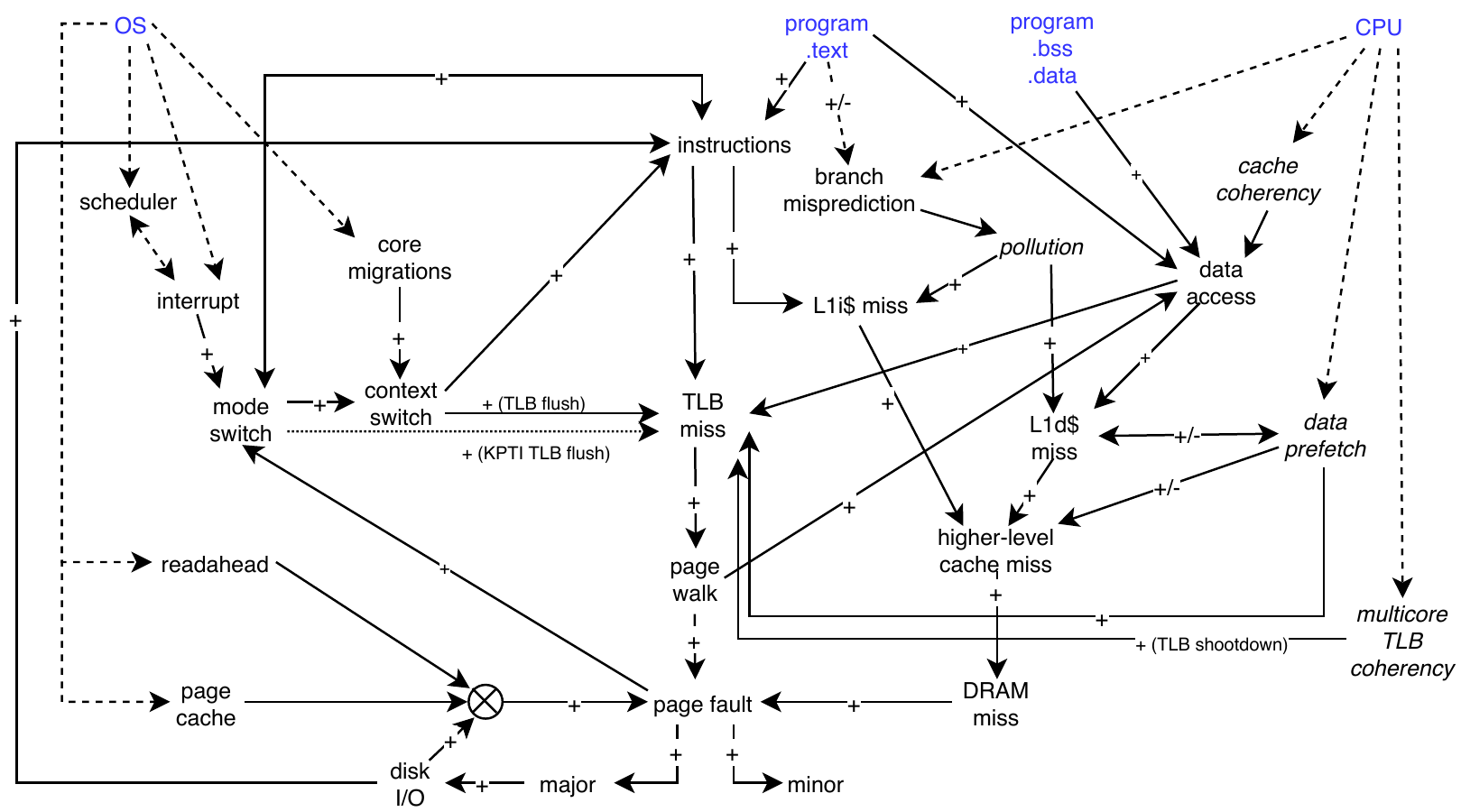}
    \caption{Event interaction map for a program running on an Intel
      Core processor on Linux. Each event itself may cause processor
      cycles, and inhibit ($-$), enable ($+$), or modulate ($\otimes$)
      others. 
      \vspace{1em}}
    \label{fig:teaser}
  \end{teaserfigure}

\maketitle


\section{Introduction}
Countless performance tests of software are available online in blogs,
articles, and so on, but often their significance can be refuted by their
measurement setup.  Speedups are usually reported in units of time,
yet without any introspection how exactly the differences came to
live. Not only are the results questionable when the software runs on a different machine,
but even on identical processors with identical memory configurations
and peripherals, there are many external factors that can influence
the result. One such factor is the operating system (OS)
itself. Different configurations and usages of OSes might nullify or
magnify some effects, and thus the performance measurements do not
necessarily reflect the characteristics of the software that we
actually want to analyze and improve.

In this report, we describe how performance measurements of user
software should be conducted, when running on a mainline Linux OS and
a modern multi-core processor. Specifically, we are concerned with
real-time measurements taken on the real hardware, providing
quantitative information like execution time, memory access times and
power consumption. 
We expect that the setup of the OS and hardware have a significant
impact on the results, and that a proper setup can greatly reduce
measurement errors.

This document is structured as follows: We start with a general survey
of microarchitectural and OS elements affecting performance in modern
processors in Section~\ref{sec:micr-os-perf}. In the next section we
list possible sources of measurement errors, followed by a proposed
measurement setup in Section~\ref{sec:experiments}. Finally, we show
several examples of how the setup influences the results, before we
conclude this report.

\section{Microarchitectural and OS Performance Modulators\label{sec:micr-os-perf}}
%
%
We start by giving an overview on hardware and OS features and how
they influence the performance of software. In essence, this section
is an elaboration of the interactions depicted in
Figure~\ref{fig:teaser}.  The well-informed reader might directly
proceed with the subsequent section, which identifies sources of
measurement errors based on the details presented here.

Although we try to keep the explanations generic, it would be
impractical to make only statements that cover any conceivable
processor.  The details are therefore given for an Intel
2\textsuperscript{nd} generation Core\textsuperscript{TM} x86-64
microarchitecture (``Sandy Bridge''). This is a pipelined, four-width
superscalar multi-core processor, with out-of-order processing,
speculative execution, a multi-level cache hierarchy, prefetching and
memory management unit (MMU). This kind of processor is currently used
in high-end consumer computers, and has well-proven architectural
features that we expect to see in embedded processors in the coming
years; many of them are already available in ARM
SoCs~\cite{ArmCortex}.  The details of our machine are summarized in
Tab.~\ref{tab:host}. As operating system (OS) we consider Linux,
specifically in an \emph{SMP} configuration.


\begin{table}[htb]
  \centering
  \footnotesize
  \begin{tabular}{ll}
    Unit & Properties\\
    \toprule
    Processor & Intel Core i7-2640M @2.8GHz, dual-core\\
    Microarchitecture & ``Sandy Bridge'', Microcode version 0x25\\
    L1i/d cache & 32kB, 8-way, 64B/line, per core\\
    L2 unified cache & 256kB, 8-way, 64B/line, writeback, per core\\
    L3 uncore cache & 4MB, 16-way, 64B/line, shared between all cores\\
    TLB & 2-level, second level unified, per core\\
    OS & Debian 8.11 GNU/Linux SMP 3.16.51-3\\
    \bottomrule
  \end{tabular}
  \caption{System specifications}
  \label{tab:host}
\end{table}

Along this report, we will give some numbers to illustrate the
magnitude of some effects. These numbers are given to the best of our
knowledge, the best available vendor documentation, and supported by
measurements that we have taken.  The numbers are all based on the
following system \emph{penalties}, caused by the typical design of
CPUs and OSes.

\noindent\textbf{Penalties:} The various latencies for our system are
summarized in Tab.~\ref{tab:penalties}, and discussed in the
following. They are taken from the processor
documentation~\cite{intelopt}, and extended by measurements using
\texttt{lmbench}~\cite{mcvoy1996lmbench} and
\texttt{pmbench}~\cite{yang2018pmbench}. The values denoted ``best
case'' stem from the manufacturer documentation. Our measurements
show that these values are also the \emph{most frequently} observed
ones.  Note, however, that penalties can (and should) be hidden by
out-of-order processing. That is, not every page walk delays
computation for 30 cycles.

\begin{table}[htb]
  \begin{minipage}{\columnwidth}   
    \small\centering
    \begin{tabular}{llr}
      \textbf{Event} & \textbf{Condition} & \textbf{Latency [cycles]}\\
      \toprule
      L1 cache hit & best case & 4\\
      L2 cache hit & best case & 12\\
      L3 cache hit & best case & 26..31\\
      DRAM access & best case & $\approx$ 200\\
      branch misprediction & most frequent & 20\\
      TLB miss & 2nd-level TLB hit & 7\\
      page walk & most frequent & $\approx$30 \\
      minor page fault & most frequent & $\approx$1,000 .. 4,000\\
      major page fault & most frequent & $\approx$260k .. 560k\\
      context switch & best case& $\approx$3,400\\
      \bottomrule
    \end{tabular}
  \end{minipage}
  \caption{Penalties for system in Table~\ref{tab:host}\label{tab:penalties}}
\end{table}

\noindent\textbf{Software Performance:} In this report, we look at software
performance, mainly from a timing point of view. Specifically, for the
execution time of the process, we only consider the time where the
processor executes instructions on behalf of the process (including
kernel code and stalls), but not sleeping or waiting states, since the
latter are either voluntary or depend on the execution context and not
the program itself.


\subsection{Microarchitecture}
\begin{figure*}[p]
  \centering
  \includegraphics[height=.93\textheight]{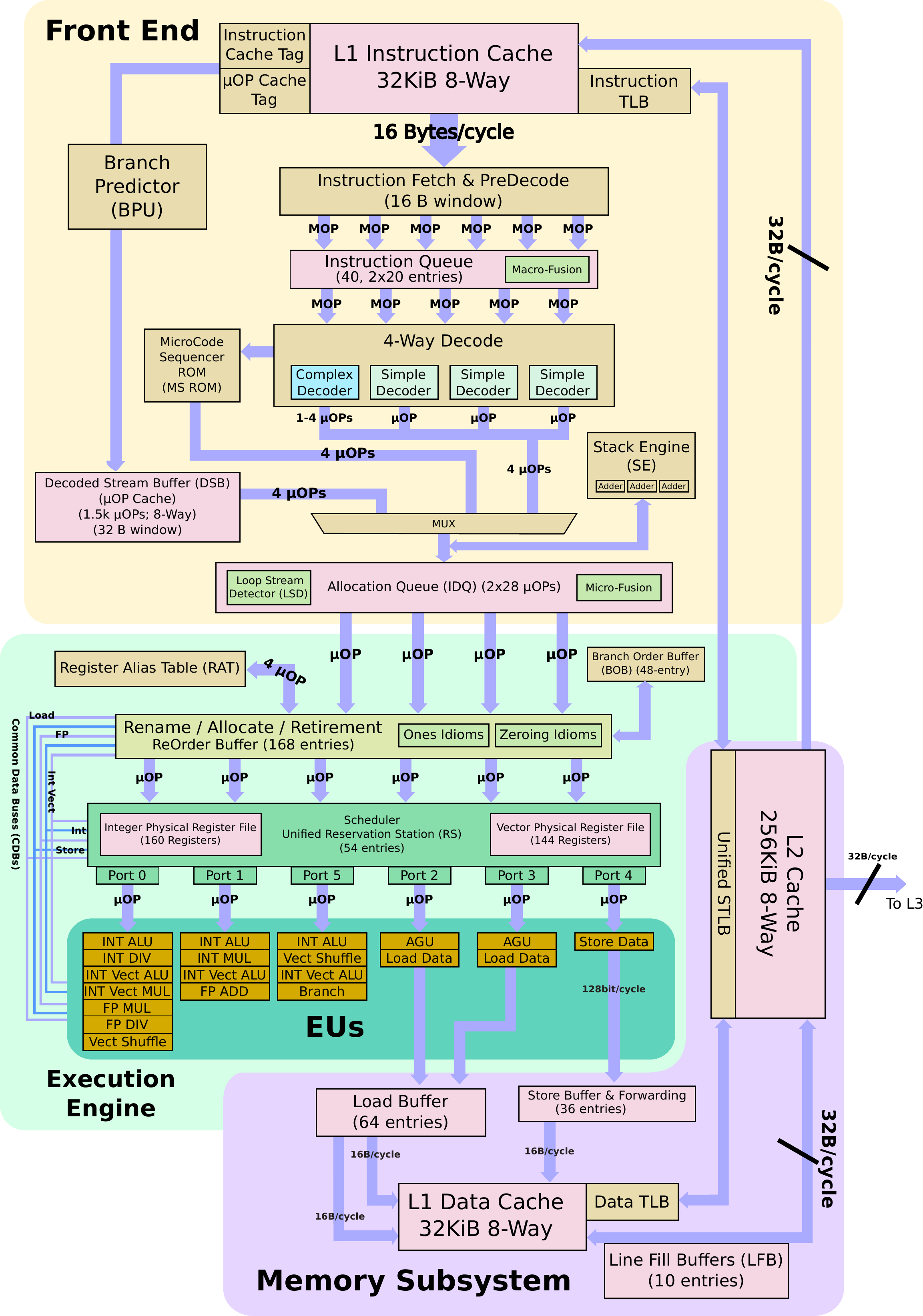}
  \caption{Microarchitecture of Intel Sandy Bridge as example of a
    superscalar out-of-order processor with caches, branch predictor and
    prefetcher. Taken from~\cite{wikichip_snb}, and checked for
    consistency against manufacturer datasheets~\cite{intelopt}.}
  \label{fig:snb}
\end{figure*}

As an overview about the features discussed next, consider the
microarchitectural block diagram shown in Fig.\ref{fig:snb}.

\subsubsection{Number of Cycles and Instructions}
As a single number indicating program speedup, we first look at the
number of processor cycles spent on execution. Lower is better.  The
number of clock cycles is primarily driven by the instructions being
executed, whereas the exact relationship is defined by the
implementation of the microarchitecture. The relationship is usually
nontrivial and not fully documented, therefore we have forgone
indicating the relative influence of events of clock cycles in
Fig.\ref{fig:teaser}. We will provide typical numbers in the following
descriptions, as far as they are known.

If we count both the number of retired instructions and processor
cycles, we can compute the ratio \emph{instructions per cycle} (IPC).
While often used as a first performance indicator, this figure is
highly program-specific, and not helpful for judging our software. For
example, a program may execute faster than before although the IPC has
dropped, simply by virtue of a reduced number of instructions.

\noindent\textbf{Micro-operations (uops):} Some processors split
instructions into multiple smaller
operations~\cite{ArmCortex,intelopt}, which provides more
possibilities for out-of-order processing~\cite{Agner2018}. On such
targets, it is more meaningful to measure uops instead of
instructions, since a single instruction may become a variable-length
series of uops. Some, usually more complex instructions, may even
invoke \emph{Microcode}.

\noindent\textbf{Microcode:} On some processors, instructions are not
hardwired but interpreted, to be broken down to hardwired machine code
during execution. This abstraction layer can incur a slowdown, but it
allows upgrading processors during their lifetime, e.g., to fix
processor errata.  On x86, microcode is only used for few complex
instructions; specifically, only for instructions generating more than
four uops in our processor~\cite{Agner2018}. The performance can be
impacted in two ways: First, switching between the regular instruction
stream and Microcode may incur a penalty, and second, Microcode
generating a lot of uops may be limited by the frontend bandwidth.

\noindent\textbf{Uops Cache:}
Instructions are decoded into uops by a hardware circuit that can be
limited in throughput, causing a bottleneck if the average instruction
length exceeds its capacity. Some processors add a cache to store
decoded uops, in an attempt to alleviate this problem~\cite{Agner2018}.



\subsubsection{Pipelining}
Instead of processing one instruction or uop after another, processing
is often is temporally overlapping to reduce execution time, called
\emph{pipelining}. For example, while one instruction is being
executed, the subsequent one can already be decoded to uops in
parallel, and meanwhile the uops from the previous instruction can be
retired. Today it is very rare to find processors or even
microcontrollers that do not implement some form of pipelining.

\subsubsection{Out-of-Order and Superscalar Processing}
Modern processors implement dynamic scheduling of
instructions~\cite{Agner2018,ArmCortex}. That is, the order in which
instructions are executed, may deviate from the original order of the
instruction stream given by the program counter. This enables the
processor to hide various processing latencies, by performing other
work while waiting for an instruction to finish. For example, an
arithmetic division can take several cycles to complete, meanwhile
successor instructions can be executed, up to the point where the
result of the division is required. Similarly, memory access latencies
can be hidden. This is one of the primary reasons, while it is
nontrivial to predict the number of clock cycles that a certain event
slows down a program.

As a side effect, such out-of-order (OoO) processing enables
\emph{superscalarity}. By having more than one instance of each
functional unit (e.g., several ALUs or FPUs), it becomes possible to
execute several instructions in parallel. Thus, OoO processors often
issue multiple instructions at the same clock cycle; for our machine,
four uops are issued simultaneously, thus the maximum IPC is four.

An algorithm for OoO was proposed by Robert Tomasulo in
1967~\cite{tomasulo1967efficient}, and today's out-of-order processor
implementations still follow the same
concept~\cite{intelopt,Agner2018}. The OoO implementation is often
called \emph{execution engine}, and for the purpose of our performance
measurements, the following constituents must be considered:
\begin{inparaenum}[1.]
\item Register Renaming and
\item Scheduler.
\end{inparaenum}

\noindent\textbf{Register Renaming}: Register names used by the compiler are
those defined by the instruction set architecture (ISA), called
\emph{architectural} registers. Due to the limited number of
registers, the compiler might be forced to re-use registers for
computations that are otherwise not related, creating false data
dependencies. However, actual implementations of the ISA may have more
\emph{physical registers} than architectural ones.  Therefore, the
first step is to map architectural to physical registers, while
resolving some false dependencies. This renaming process
therefore improves superscalar and OoO processing~\cite{Agner2018}.

\noindent\textbf{Scheduler}: The renamed operations now enter the main core of
OoO processing, the \emph{scheduler}. Here operations are queued up
for execution units in \emph{reservation stations}, and will be held
there until all operands become available. When this becomes the case,
the operation is started on the execution unit. As soon as it
completes, results are propagated to all subscribers (e.g., registers
or reservation station entries waiting for a result), and the
operation is forwarded from the reservation station to the Reorder
Buffer. From here, most schedulers also take care of retiring the
instructions in their original order. A bottleneck may occur when the
scheduler stalls execution due to lack of free reservation stations.

\subsubsection{Branch Mispredictions}
Many microarchitectures perform some kind of branch prediction, to
hide the latency for loading the instructions and data after a
branch~\cite{Agner2018,ArmCortex}. Predictions for both the outcome of
two-way branches, and the (possibly multi-way) target address of
indirect branches, are being made by the branch prediction unit.  

If the branch is incorrectly predicted, then the pipeline and other
resources must be flushed, which means there is a time penalty to
fetch and decode the correct instructions.  The magnitude of the
penalty primarily depends on the pipeline depth. In case of Sandy
Bridge, the penalty for flushing and resuming execution is about 20
cycles~\cite{intelopt,Agner2018}.


\subsubsection{Speculative Execution}
The time window between branch prediction and learning the actual
branch outcome is spent with \emph{speculative execution}. That is,
the processor continues the control flow at the assumed branch target,
and buffers the results until the actual branch target becomes known.
Whenever the prediction was incorrect, it flushes the pipeline as
described above. If the prediction was correct, the speculatively
executed instructions are allowed to be released from the buffer
(``retire''), and no time was lost waiting for the branch
outcome. 

There is one lesser known side effect, however, dubbed
\emph{pollution} in Fig.\ref{fig:teaser}. Although instructions
executed during mis-speculation are not retired, they can still cause
changes in cache and buffer states. These effects are \emph{indirect}
cost of branch mispredictions, which manifest themselves during later
execution. These effects have recently been exploited in the \emph{Spectre}
and \emph{Meltdown} vulnerabilities~\cite{kocher2018spectre}.

Last but not least, even perfect speculation might become a
performance bottleneck in some cases. All speculatively taken branches
are stored in the \emph{branch order buffer} (BOB) until they are
confirmed. However, too many speculations in a short time window might
cause the BOB to fill up, which in turn stops the issue of new
uops~\cite{intelopt}.

\subsubsection{Machine Clears}
When multiple threads can run truly in parallel (as on SMP systems and
especially with OoO processing), the ordering of memory accesses must
be monitored and ensured. If the CPU detects that accesses complete
differently to program order, a \emph{machine clear} is
performed. This entails undoing some operations, flushing the
pipeline, and re-starting with the correct
operands~\cite{intelopt}. The cost is comparable to a branch
misprediction.  Further causes for machine
clears are self-modifying code and illegal AVX addresses.

\subsubsection{Cache Hierarchy and Misses}
Although memory access is fast these days, it can still be orders of
magnitudes slower than the processor, which has become known as the
\emph{Memory Wall}~\cite{DBLP:journals/sigarch/WulfM95}. Consider the
penalty for a major page fault (disk access) shown in
Tab.~\ref{tab:penalties}: At least five orders of magnitude lie
between the processor and the disk access, although we use a
relatively fast Solid State Disk (SSD).  \emph{Caching} is the
omnipresent approach to counteract this issue, by preloading or
latching all data in faster, smaller memories, and exploiting the
\emph{principle of locality}. That is, data that has been recently
accessed, is likely to be accessed soon in the future again. It thus
makes sense to buffer recently used data in caches. Every time a data
item is requested (``data access'' in Fig.\ref{fig:teaser}), we check
for the desired data in the faster cache, before accessing the slow
memory. If we find the data, called a \emph{cache hit}, we have
circumvented waiting for the slow memory. If the cache does not
contain the data, called a \emph{cache miss}, we have to pay the
penalty for accessing the slow memory (usually DRAM). After this, the
data is placed in the cache for future reference.

Since caches have to be small to be fast, there are inevitably
situations when data is not in the cache, and needs to be loaded from
slower memory. Even if we were able to perfectly predict what data
(including instructions) is needed, then there are still compulsory
misses on first access.  As a result, execution can be slowed down by
one or two orders of magnitude even with fast caches and very high hit
ratio.  For example, let us assume each instruction takes one cycle,
and that each cache miss costs five extra cycles. Then, even with
$p=95\%$ hit ratio, a program with $X$ instructions would take
$X+X(1-p)5=1.25X$ cycles, i.e., experience a 25\% slowdown. Measuring
cache behavior is therefore important.

\noindent\textbf{Victim Caches:} These are small and fully associative caches,
holding items evicted from a larger, not fully\--asso\-ciative
one. Each miss in the large cache is first looked up in the victim
cache, before the slower memory is consulted. This masks miss times
for temporally close conflict misses. From a practical point of view,
the victim cache need not be considered separately; instead, a victim
hit can be seen as a hit in the faster cache. Vice versa, a victim
miss can be seen as a hit in the next-slower memory or cache.

\noindent\textbf{Hierarchy:} As most modern processors, our machine uses a
\emph{hierarchy} of caches. That is, there are three caches in a
cascade (three ``levels'') before the slower DRAM
is accessed.  The first level (L1) is the fastest/smallest and
separate for data (L1d) and instruction (L1i), see
Tables~\ref{tab:host} and \ref{tab:penalties}. Leaving aside special
architectural tricks, this is the only level that can be accessed
directly by the CPU. Higher levels (L2, L3) are larger and slower, and 
usually unified. Depending on the processor, some caches can be 
 shared with peripherals or other processors.

\subsubsection{DRAM Access}
The next-larger memory after the cache hierarchy, is Direct Random
Access Memory (DRAM). Conceptually, the cache hierarchy acts as a
frontend and buffer to DRAM accesses.  Only if the lookup of data or
instructions missed at all levels in the cache hierarchy
(Fig.\ref{fig:teaser} ``higher-level cache miss''; not necessarily
sequentially, though), then the DRAM is consulted.  Accesses to DRAM
are typically 100 times slower than the CPU, thus the penalty missing
the entire cache hierarchy becomes steep. Even worse, DRAM is often
accessed via the \emph{Northbridge}, possibly suffering contention
with other CPUs and DMA transfers~\cite{drepper2007every}.


\subsubsection{Hardware Data Prefetching}
To further reduce access times to slow memory, many processors have a
\emph{data prefetcher} circuit, which predicts future data accesses
and actively pre-loads the data into caches. Most prefetchers are
triggered by certain access patterns in cache
misses~\cite{drepper2007every}. Some newer processors may even cross
page boundaries and thus influence the TLB~\cite[\textsection 2.4.7]{intelopt}.  In summary,
prefetch events both depend on and influence L1d cache events, as
shown in Fig.\ref{fig:teaser}.

\subsubsection{ISA Extensions, Streaming/Vector/SIMD}
Processors may extend the ISA with instructions applying the same
operation on multiple data items (vectors, single-instruction-multiple
data). Examples are the extensions AVX, SSE, MMX on Intel and AMD, and
NEON on ARM processors. Using these can greatly speed up certain
calculations, but switching to these modes may also cause extra
penalties~\cite[\textsection 9.1.2]{Agner2018}. In general, any switch
between ISA modes or extensions may cause extra penalties.

\subsubsection{Direct Memory Access (DMA)}
Accessing secondary storage, such as hard drives, but also traffic
from network cards, usually takes place via DMA, which enables
peripherals to exchange data directly with the DRAM. As mentioned
earlier, this might lead to bus contention and slow down DRAM accesses
for the processor.

\subsubsection{Cache Coherency Protocols and ring bus}
On multi-core systems, further cache accesses are caused by cache
coherency protocols, e.g., between the caches of the different
cores~\cite{drepper2007every}.  It becomes active during core
migrations, but also in the presence of data shared between cores.

\subsubsection{Neglected Features}
There are many peculiarities to each microarchitecture. We have
omitted many of such details, in an attempt to focus on those features
prevalent in most processors. Of course, these details can be
important for the performance, and a careful study of the
microarchitecture is required to see which need to be
measured. Omissions include instruction and uop fusion, loopback
buffers, register stalls, exhaustion of execution ports, cache bank
conflicts, misaligned memory access,
and store forwarding stalls. These are not considered to have a large
or systematic performance impact on Sandy Bridge~\cite{Agner2018}.

\subsection{Microarchitecture and OS Interaction}

\subsubsection{Virtual Memory and Paging}
Virtual Addressing is common in larger general-purpose and application
processors, and is done for a variety of reasons, chief among which
are process isolation and provision of a contiguous address space from
the process' point of view. However, it also enables \emph{paging},
which helps to significantly mitigate the latency of accessing slow
secondary storage, such as hard drives.  Unfortunately, Virtual
Addressing comes with a performance penalty that can vary
highly. Translation needs to be performed for every instruction and
data reference (see Fig.\ref{fig:teaser}), thus there is an
obvious incentive to minimize delays. Therefore, in practice the
address translation process is very intricate and specific to the
microarchitecture. An in-depth description can be found in~\cite{drepper2007every}.

Usually a hardware unit called \emph{Memory Management Unit} (MMU) is
responisble for the address translation.  To ensure translations do
not stall execution, most MMUs have their own caches, called
\emph{Translation-look-aside buffers} (TLBs), which hold the most
recently translated addresses. In case the buffers do not provide the
needed translation (TLB miss), then a slow search in memory has to be
conducted, called a \emph{page walk}, see Fig.\ref{fig:teaser}. On x86
machines, this means that a dedicated hardware circuit starts looking
for page table entries in memory, which incurs a penalty depending on
memory access latency. For other architectures, like some PowerPC,
this might be done in software, and thus is even slower. In case the
information was found in the page table, the TLB is updated and
execution resumes. Otherwise, the operating system is signaled a
\emph{page fault}, and has to decide whether the access is allowed.
If so, the OS updates the page table and TLB with the requested translation, and
possibly bring missing data to the DRAM. The cost for virtual
address translation therefore is consisting of cycles spent in TLB
lookup (hardware), plus page walk cycles (often hardware), plus cycles
to handle page faults (OS).

\noindent\textbf{Translation-Lookaside-Buffer:} 
TLBs are regularly flushed by the OS, since it is responsible for the
coherency between TLBs and page tables in memory, and that the TLBs do
not hold stale translations from a formerly running process. They can
either be flushed or selectively updated, depending on the OS and
hardware capabilities. Accesses after a flush are TLB misses,
therefore flushes degrade performance.
In the x86 architecture, there furthermore exists the case of
\emph{TLB shootdown}. This stems from the need to have consistent TLB
entries between multiple cores in the presence of sharing.  Since x86
does not have a coherency protocol in place, TLB contradictions
between cores are avoided by triggering flushing in hardware. Last but
not least, TLB and cache access can be executed in parallel, to reduce
latency.

\noindent\textbf{Page Walks:} Examining the page tables in memory
incurs a penalty that depends on whether the tables are cached (L1d,
L2, L3), or whether the slow DRAM needs to be consulted.  On our
machine, a typical page walk costs between 20 and 60 cycles. In an
extreme case (``TLB trashing''), this could happen for every
instruction, and thus become very expensive.  On some
microarchitectures, as in Intel Sandy Bridge, page walks go through
the caching hierarchy.
That is, page tables are buffered in L1d and below, and thus caches
can be modified by page walks. Consequently, caching behavior and TLB
misses cannot always be separated, even in the absence of page faults.
Additionally, it has recently been disclosed that Intel processors
speculatively work on possibly invalid cache entries in parallel to
page walks (see ``L1TF'' vulnerability~\cite{l1tf}). It can therefore be assumed
that even the latency of page walks is partially hidden.

\noindent\textbf{Page Faults:}
If a translation cannot be found in the page table, a \emph{page
  fault} is signalled from the MMU to the OS (Fig.\ref{fig:teaser}
bottom). There are three fundamental types of page faults:
\begin{inparaenum}
\item Invalid page fault,
\item major page fault, or
\item minor page fault.
\end{inparaenum}

\emph{Invalid page faults} are those caused by an attempt to access
addresses that are beyond the process' address space, or where
privileges are insufficient. An example are segmentation faults. We do
not discuss them further, since they are pathological events pointing
to faulty software.

%
%

\emph{Major page faults} require disk access, which is orders of
magnitude slower than the effects we are trying to observe here. For
our SSD-equipped machine, we used the
\texttt{pmbench}~\cite{yang2018pmbench} tool and found the most
frequent latency for major faults as between 262 thousand and 524
thousand cycles (coinciding with the median), same for both read and
write accesses. Additionally, the distribution is tail-heavy, similar
as in~\cite{yang2018pmbench}. That is, the estimated average is about
one million cycles, due to some accesses exceeding several dozens of
milliseconds.  \\Closely related to page faults, Linux implements a
\emph{page cache:} the virtual memory buffers data blocks of recently
used files. When files are read (e.g., via \texttt{fread} or
\texttt{mmap}), then access is by default buffered via the page
cache. Therefore, if a program is executed a second time and there was
sufficient memory, there ideally are no major page faults due to file
access. To further reduce first-time access latency, the Linux kernel
proactively reads file data from disk before it is demanded (``read
ahead''), which is no longer counted as major page fault.
Finally, if DRAM is exhausted and \emph{swapping} is enabled, unused
pages are temporarily written to secondary storage (``swapped
out''). When they are needed again, they have to be brought back to DRAM,
which also causes major page faults~\cite{Gorman}.

\emph{Minor page faults} are caused by memory allocation without disk
access, but may still be problematic for our measurements. The
memory can either be immediately allocated, or only reserved
(``lazy''). In the latter case, which is the default case in Linux, the page is
only created when the first write occurs (``copy-on-write'').  This
means that the penalty of minor page faults consists of two parts: the
cost for the fault handler itself, and the conditional copy-on-write
cost. Measuring the cost of a minor page fault is unequally more
complicated than major faults.  First, the latency is relatively
short, resulting in inaccuracy when we try to measure it using software.
Tracing kernel functions is one option that we have exercised, but it
has some non-negligible overhead in our kernel version, only giving us
an upper bound. Hardware measurements are not possible in this
case. The result is a somewhat wide range for the minor page fault
cost. Using \texttt{pmbench},
we found that the mode lies between 1,024 and 4,096 cycles, again
coinciding with the median.  Note that this includes lazy allocations.
The numbers are consistent with recent measurements of Torvalds on the successor
microarchitecture \emph{Haswell}~\cite{torvaldsminor}. 



\subsubsection{Context Switch}
Switching between processes happens frequently (discussed later in
Section~\ref{sec:scheduler-ticks}), and is an expensive operation that
influences the TLB and caches, see Fig.\ref{fig:teaser}.

Besides executing a number of instructions to save and load hardware
registers, stack pointer and PC of suspended and waking processes, the
pipeline is flushed, and the TLB must also be
updated~\cite{li2007quantifying}. On most Linux versions the TLB
update takes the form of a flush, accompanied by switching the page
table pointer. This can only be avoided with hardware that supports
process context identifiers (PCIDs) and with newer kernels supporting
this feature -- such as x86 on Kernel 4.14 onwards~\cite{PCID}. The
penalty for such TLB invalidation is called the \emph{direct cost} of
a context switch~\cite{li2007quantifying}.

However, the CPU caches are also affected, which incurs \emph{indirect
  costs}: Because processes share caches among themselves and also with
the kernel, the waking process may not find its data in the caches as
had been left when it got suspended, and experience cache misses. This
effect magnifies with growing working set
size~\cite{li2007quantifying}.  Context switches can happen
involuntarily due to interrupts, or voluntarily due to system calls
(both discussed later). Using \texttt{lmbench}, we found that context
switches on our system take at least 3,400 cycles, with a frequent
value around 30,000 cycles. Note that this number would increase if a
core migration happens at the same time, which is explained next.

\subsubsection{Core Migrations and Load Balancing}
The OS (and the hardware, if Hyperthreading is enabled) may migrate a
running process between cores (``core migrations'' in
Fig.\ref{fig:teaser}), to balance load or thermal stress. This causes
a context switch with additional overhead. The process must be
stopped, copied to another core's run queue, cache lines are moved to
the new core~\cite{drepper2007every}, and only then both scheduling
domains are released. Naturally, this requires to run a lot of kernel
code, which in turn increases chances for events like branch
mispredictions, cache pollution and page walks.  We have observed
latencies of beyond 100,000 cycles for migrations, depending on the
working set size.

\subsubsection{Mode Switch\label{sec:mode-switch}}
Switching between user and kernel mode is not a context switch, and
thus has lower overhead. These switches are caused by system calls
done by the running program (thus the bidirectional interaction
between instructions and mode switches in Fig.\ref{fig:teaser}) for
which the kernel is supposed to perform some work on behalf of the
calling process. The kernel is then said to be in \emph{process
  context}.

In Linux on x86, these calls involve copying the arguments to
registers, triggering a trap (CPU changes to kernel mode), executing
the trap handler (which copies the arguments from the registers to the
kernel stack and then performs its work), and eventually returning to
user mode. Depending on the processor and OS, the TLB might be
invalidated. This is the case for x86 since the \emph{Spectre} and
\emph{Meltdown} vulnerabilities in 2017, where now \emph{kernel page
  table isolation} (KPTI) invalidates entries during each mode switch
to protect the kernel from attacks, unless the hardware supports PCIDs
(see above). Finally, when exiting the kernel mode, a context switch
to a different process might take place.

Our kernel has KPTI enabled, but no PCID support. We created a test
program to measure the direct penalty from KPTI.  We found that the
fastest syscall drops from at most 660 cycles down to less than 130
cycles, when KPTI is disabled\footnote{This upper bound was measured
  as the minimum latency over many calls of \texttt{getuid}, with the
  program given in Appendix~\ref{sec:mode-switch-benchm}. This program
  had experienced a 3x speedup!}. Kernel developers have measured
large slowdowns as well, with up to 30\% in networking
code~\cite{KAISER}. That is significant, and thus needs to be
considered during measurements.

In summary, mode switches also have a direct penalty caused by the
call overhead, and indirect penalties caused by TLB and cache effects,
depending on kernel version and processor. 

\subsubsection{Interrupts} 
Typically, a few dozens of interrupts\footnote{see
  \texttt{/proc/interrupts}} can arrive at any point in time. For
example, there are \emph{thermal interrupts} in case the hardware
overheats, and \emph{machine check exceptions} indicating hardware
errors in the CPU. Some interrupts cannot be avoided, while others can
be disabled or redirected to different cores. Interrupts do not
directly cause context switches as described above. Instead, the CPU
itself saves and restores very few registers (notably, the PC), and
then a mode switch happens~\cite{kfund}, see also
Fig.\ref{fig:teaser}. After this, the interrupt handler must take care
of the rest. Specifically, the kernel then is running in
\emph{interrupt context}, as opposed to process context, and the work
being done here is not attributed to any user process.

The interrupt handler itself is supposed to be fast and must not
suspend execution. This is also called the \emph{top half}. Interrupts
which require longer-lasting or I/O work to be done, must defer such
work for later processing in a so-called \emph{bottom
  half}~\cite{ksoftirq}.

Once the top half handler finishes, the OS switches back to user mode. As
explained before, a context switch might happen during this
transition. Specifically, if the interrupt had deferred some work to a
bottom half, then a context switch to a kernel thread might happen at
this point in time, to finish the interrupt work.

One special periodic interrupt is the \emph{local timer interrupt}
driving the scheduler, also called ``scheduling clock ticks'',
and discussed next.

\subsubsection{Scheduler Ticks\label{sec:scheduler-ticks}} 
By default, the Linux scheduler runs periodically, to select which
process is executed on the hardware for the next time slice. This is
achieved by a timer interrupt firing typically every four
milliseconds. Considering that interrupts cause mode switches and
possibly context switches (see Fig.\ref{fig:teaser}), running the
scheduler itself may unnecessarily suspend our user task. Luckily,
there are two options to configure the kernel differently, commonly
referred to as ``tickless''~\cite{tickless}:

\begin{enumerate}
\item \emph{Dyntick-Idle}, enabled by the kernel configuration option
  \texttt{CONFIG\_NO\_HZ\_IDLE=y}, describes a kernel configuration
  that omits scheduling ticks when there is no task to be
  executed. While this is often the default for desktop and embedded
  systems to save energy, but is not useful for our purpose.
\item \emph{Adaptive ticks}, enabled by kernel configuration option
  \texttt{CONFIG\_NO\_HZ\_FULL=y}, describes a kernel configuration
  that additionally turns off ticks when there is only one runnable
  task, thus preventing unnecessary interruptions. On top of the
  configuration option, it is necessary to add a kernel boot parameter
  specifying for which CPUs it shall be applied. RCU calls need to be
  offloaded to other CPUs.
\end{enumerate}
Both of these configurations have disadvantages, as well, including
increased number of instructions to switch to/from idle mode, and more
expensive mode switches~\cite{tickless}. Using again the program shown
in Appendix~\ref{sec:mode-switch-benchm}, we have found that mode
switch times in our system increase to about 700 cycles with adaptive
ticks and KPTI disabled. Consequently, tickless operation may not be
beneficial for workloads with many syscalls, or for those going
idle frequently.

\subsubsection{Neglected Features}
The most important mechanisms of OS-hardware interaction have been
explained. Nevertheless, there are many other features and effects
that depend on the specific version of OS. Some omissions
include speculative paging~\cite{spec_page}, as that is a recent
development not commonly used, yet, and page migration between NUMA
nodes, since our target is a single NUMA node.

\section{Measurement Errors\label{sec:jitter}}
This section summarizes features that can create measurement errors,
building on the explanations given before and visualized in
Fig.\ref{fig:teaser}.  By \emph{error} we subsume all effects that
stem from sources other than the software under analysis and, when not
properly controlled, would therefore mislead us in a systematic or
random direction.  For example, we consider effects caused by other
processes running in the background of an operating system as
measurement error. Inter alia, they share caches with the software
under analysis, and thus the caching behavior of our software can look
significantly worse if there exists a data-heavy background process.

Table~\ref{tab:errors} depicts all major sources of measurement
errors, together with a classification whether they are caused by
hardware (HW) or software (SW), their measurable impact, and the time
when the effects manifest in the measurements. Additional explanations
are given in the notes below.

\begin{table*}[htbp]
  \centering
  \begin{tabularx}{\textwidth}{llXll}
    \toprule \textbf{Source} & \textbf{Class} & \textbf{Impact} &
    \textbf{When} & \textbf{Note}\\\midrule
    Speed Stepping/Frequency Scaling & HW & varying execution time & immediate &\\
    TLB Shootdown & HW & slower address translation & lagging &\\
    DMA Transfers & HW/SW & slower memory access & immediate & (\ref{note:dma})\\
    Cache Coherency* & HW  & additional accesses to cache & immediate \& lagging &(\ref{note:cc})\\
    Hardware Prefetcher* & HW & more cache \& TLB accesses, less or more misses & immediate \& lagging & (\ref{note:pf})\\
    Load Balancing/Migrations & SW & more \& TLB accesses, less or more misses  & immediate \& lagging\\
    Interrupts & HW/SW & longer execution time, more mode switches & immediate \& lagging & (\ref{note:irq})\\ 
    Mode Switches & SW & more cache/TLB misses, more context switches & immediate \& lagging & (\ref{note:ms})\\
    Context Switches & SW & more cache/TLB misses, longer execution time & immediate \& lagging & \\
    Allocator & SW & more context switches, different cache usage & immediate \& lagging & (\ref{note:alloc})\\
    Scheduler & SW & more or less context switches & immediate \& lagging & (\ref{note:sched})\\
    Major Page Fault* & SW/HW & more context switches & immediate \& lagging & (\ref{note:major})\\
    Peripherals & HW & more variance in memory access, cache misses & immediate \& lagging & (\ref{note:periph})\\
\bottomrule
  \end{tabularx}
  \caption{Summary of sources for measurement errors. Sources marked with (*) may or may not be considered creating measurement errors, depending on the intended observation.}
  \label{tab:errors}
\end{table*}

\noindent\textbf{Notes for Table~\ref{tab:errors}:}
\begin{enumerate}[(a)]
\item\label{note:cc}Only traffic not caused by the software under
  analysis is considered an error. Such traffic might be caused by all
  other processes running on the same core, including the kernel. See
  process isolation.
\item \label{note:pf}If the number of demand data accesses shall be
  measured, then the prefetcher skews the results, otherwise it is not
  considered a source of error. Also, prefetching has the same effect
  as demand access on caches and TLB, thus it can skew TLB access
  metrics.
\item \label{note:irq}Only interrupts not caused and required by the
  software under analysis are considered source of errors. Interrupts
  may cause other processes to be scheduled to execute bottom halves.
\item \label{note:ms}The mode switches themselves are often required
  to execute system calls. But mode switch overhead can be highly
  different depending on OS and HW. Therefore, it may or may not
  contribute a significant error.
\item \label{note:alloc}Allocators for virtual memory have different
  characteristics. Some may consume significantly more/less memory
  than others, and some may force context switches.
\item \label{note:sched}Non-tickless kernels may generate unnecessary
  context switches and impair TLB, caches and thus performance. None
  of those switches are caused by the software under analysis, and
  thus considered errors.
\item \label{note:major}Major page faults cause disk access. If the
  effects to be measured are rather small compared to disk access
  times, then major faults would mask those effects due to large
  jitter, and must be avoided. Furthermore, it is worth noting that a
  process accessing a file might be sent into waiting state, and then
  the number of unhalted processor cycles is not
  incremented. Consequently, I/O is not directly visible in
  counter-based measurements, and even I/O-limited applications can
  still exhibit a high IPC. Indirectly, however, it can be seen that
  the process spends less CPU time than wall-clock time, pointing to
  an I/O-bottleneck.
\item \label{note:dma}Depending on the microarchitecture, DMA might
  slow down DRAM access because of increased traffic on the
  Northbridge~\cite{drepper2007every}.  Naturally, DMA also generates
  interrupts. Furthermore, cache coherency might be a
  concern~\cite{kernelDMA}, and generate extra cache traffic.
\item \label{note:periph}Some caches might be shared with peripherals
  and thus be influenced by their operation. For example, the GPU and
  the processor share the L3 cache on our machine. This means that
  even different monitor setups can change cache miss statistics and
  bandwidth benchmarks\footnote{We have observed a 10\% decline in
    throughput in the STREAM benchmark~\cite{mccalpin1995sustainable},
    when executed on a dual-monitor vs. single-monitor setup}.
\end{enumerate}

\subsection{Short-Living Programs}
When an OS is used, there is necessarily an overhead for starting and
terminating a process. This can become significant when we measure the
performance of programs that have a short execution time. After all,
most developers are not trying to optimize the OS or its libraries,
but their own software. To give a rough number, programs that execute
less than a million instructions should be considered to be too short,
with the specific number being subject to the amount of work the OS
has to do.

\begin{figure*}[p]
  \centering
\subfloat[]{
    \includegraphics[angle=90,height=.91\textheight]{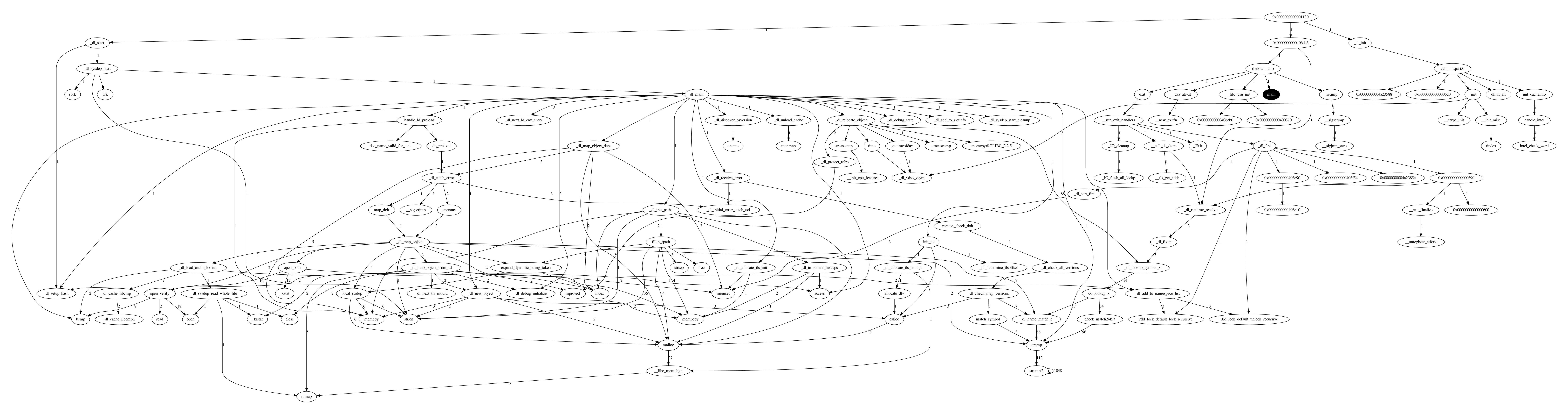}
}
\subfloat[]{
    \includegraphics[angle=90,height=.91\textheight]{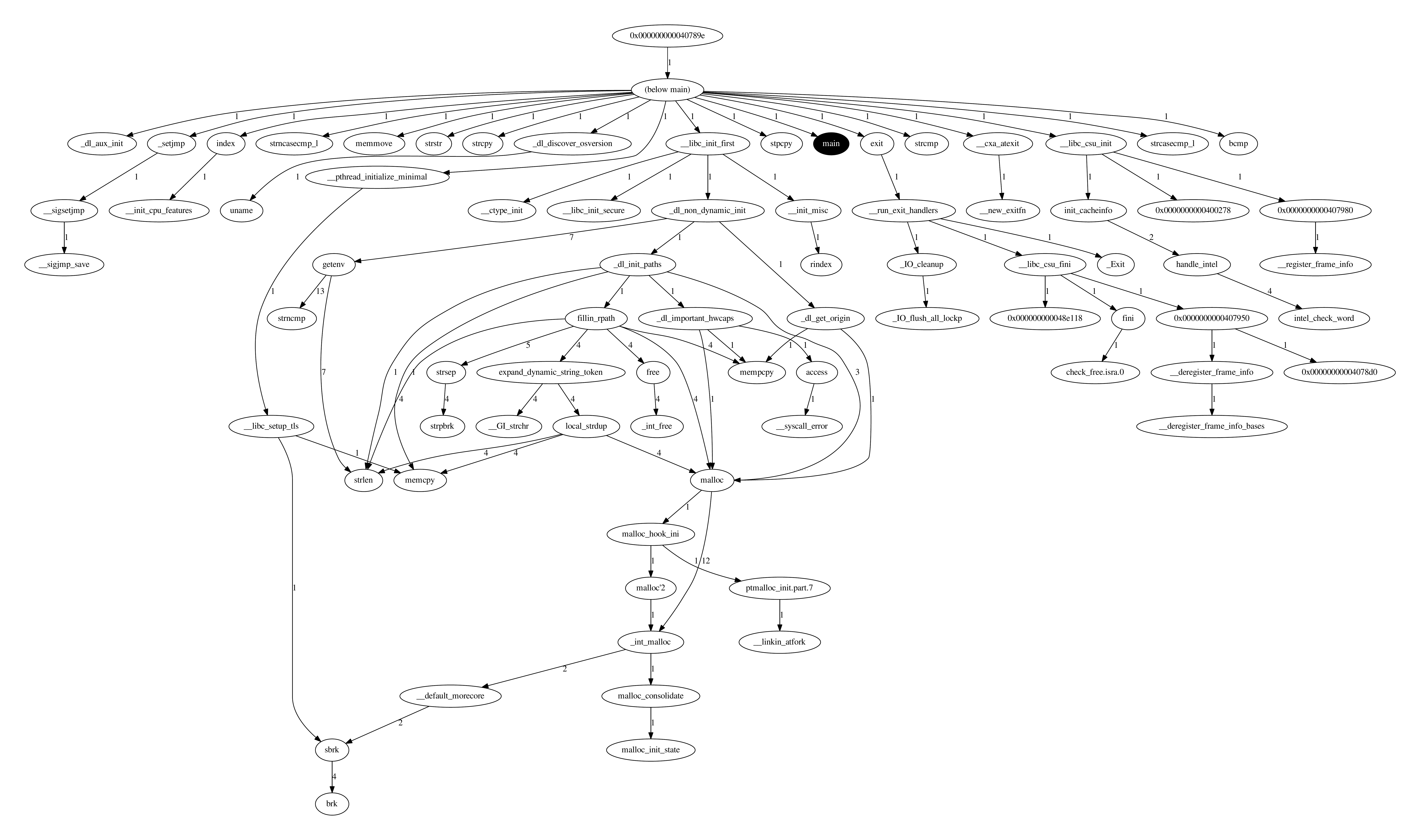}
}
\caption{Call graphs of a short-living program, suggesting that most of
  the work being done is OS ``overhead''. The user function is
  highlighted in black. Both figures show the same program, where (a)
  is dynamically linked and (b) statically linked.}
  \label{fig:callg}
\end{figure*}

As an example, consider the call graph of a program (the
\texttt{nsichneu} from the M\"alardalen WCET benchmarks, compiled with
gcc and \texttt{-O3}) shown in Figure~\ref{fig:callg}a. The user code,
all in a single function (black node in the upper right half),
executes about 40,000 instructions. The entire program, from startup
to after termination, requires about 150,000 instructions. The OS
performs tasks such as \emph{dynamic loading} (starts in upper left
corner and spreads about two thirds into the picture), which locates
and loads into memory all dynamic libraries used by the program. The
actual user code is executed thereafter, and followed by another
cascade of calls performing cleanup actions. Figure~\ref{fig:callg}b
shows the same program with static linking. The remaining overhead is
mainly from the C library, although this programs does not make any
explicit library calls, as evident from the call graph. In this
example, the user code is only responsible for about 26\% of the
processor cycles with dynamic linking, and for 75\% with static linking.

In summary, if we are not aware of such OS and library overhead when
looking at the measurement results, we may draw conclusions that
merely reflect the operating system and its libraries, rather than the
software that we are trying to analyze and improve.

\section{Measurement Setup}\label{sec:experiments}
%
%

In this section, we propose a measurement setup on SMP Linux, that
minimizes measurement errors and thus provides a faithful
quantification of the software under analysis, while suppressing other
influences as far as
possible.

All event interactions described in previous sections are highly
depending on the microarchitecture, the operating system and its
configuration, next to the program under analysis itself.  We describe
our measurement setup for the system detailed in Table~\ref{tab:host},
as a representative of an advanced superscalar out-of-order processor
with operating system, MMU, caches, prefetchers and so on.  For
different targets or OSs, the setup has to be adapted accordingly.

The main focus of our setup is how to control variables that would
otherwise lead to the measurement errors listed in the previous
section. Therefore, many of the suggestions given here are about
parameters and configuration of both the OS and the CPU. In the end,
we briefly describe the measurement process itself.

\subsection{Control Variables}

\subsubsection{Isolating and Pinning the Process}
First, one or more (yet not all) CPU cores were isolated from the
scheduler and SMP balancing, which prevents other userspace tasks from
interfering (kernel boot parameter \texttt{isolcpus}). We recommend
to leave out CPU0, since one core is required to process the offloaded
tasks, and CPU0 usually serves interrupts that cannot be moved to
other cores (e.g., some DMA controllers). 
Additionally, Hyperthreading was turned off in BIOS, to avoid hardware
context switches, same-core migrations and some known errata with the
hardware counters. The process under analysis was subsequently pinned
to the isolated cores with command line tool \texttt{taskset}. Note
that each subprocess/thread needs pinning if a range of cores is
isolated, since otherwise migrations might happen.

\noindent\textbf{Interrupts:} The affinity of interrupts was set to the
non-isolated cores, preventing them from skewing the
measurements\footnote{\texttt{/proc/irq/default\_smp\_affinity}}.
\emph{Thermal interrupts} were prevented by throttling the CPU speed
such that full load does not lead to critical temperatures. This can
be done by setting the governor's limits in \emph{sysfs} or tools like
\texttt{cpufreq}.  \emph{Machine check exceptions} were disabled by
kernel boot parameter \texttt{mce=off}.

\noindent\textbf{Ensuring tickless operation:} To minimize the number of
context switches, a kernel with adaptive ticks should be used, and
enabled with kernel boot parameter \texttt{nohz\_full}.  The remaining
challenge is to keep kernel threads off the run queue, to prevent scheduler
ticks from getting generated. There are several factors that can cause
runnable kernel threads, which are discussed in~\cite{kthreads}.  One
such factor are RCU calls, which should be disabled for the isolated
CPUs with kernel boot parameter \texttt{rcu\_nocbs}.  Another reason
are bottom halves of interrupts as discussed earlier, which can be
prevented by setting the interrupt affinity. 

\noindent\textbf{Allocator Selection:} Yet another reason why tickless
operation fails, may be the kernel's memory allocator. Depending on
the distribution, different algorithms (called ``SLAB'', ``SLOB'' or
``SLUB'') can be in use, with different impact on cache miss/hit
metrics and execution time. Notably, the SLAB allocator requires
periodic cleanups, which enables a kernel process and prevents
tickless operation. 
%
The best option is the newest, and in Desktop distributions most
commonly used allocator, the SLUB allocator (one exception is Debian,
using SLAB and thus rendering adaptive ticks ineffective). Changing
the allocator requires building a custom kernel.

\noindent\textbf{Watchdogs:} More context switches could be caused by the
watchdog. This can be prevented with kernel boot parameters
\texttt{nowatchdog} and \texttt{nosoftlockup}.

\noindent\textbf{Real-Time Priority:} Depending on the Linux version, some
kernel threads~\cite{kthreads} are still active on isolated cores and
can cause unwanted context switches (e.g., \emph{kworker}). To avoid
this, we perform our measurements at real-time priority using the
command line tool \texttt{chrt}. Additionally, real-time throttling
must be disabled (by setting \texttt{sched\_rt\_runtime\_us} in
\emph{procfs}), since otherwise some Linux versions force one
involuntary context switch per second for CPU-bound tasks.


\vspace{1em}\noindent\textbf{Summary:} The show configuration results
in no other user processes running on the isolated cores, and in no
CPU migrations taking place of our process(es). Some interrupts may
still be left, but these should only cause mode switches. This can be
verified using \texttt{perf}\footnote{\texttt{perf record -e
    "sched:sched\_switch"} ...}. 

\subsubsection{Fixing Processor Speed}
If the processor supports speed stepping/frequency scaling, then the
clock speed of the processor should be fixed for two reasons. First,
if the absolute execution time is part of the measured metrics, a non-fixed
clock speed could yield different results in subsequent runs,
depending on how the governor selects the frequency. Second, we
conservatively prevent a potential impact of frequency scaling on
processor cycles; usually there is not enough information in the data
sheets to conclude there is no influence.

Further, fixing the clock speed also ensures more stable DRAM access
times when measured in processor clock cycles. DRAM runs at its own
bus frequency, which means that a higher CPU frequency will make DRAM
bottlenecks stand out more. This is important for backend-bound
benchmarks, where the bottleneck becomes more severe with increasing
processor frequency.

Fixing the processor speed is best done in BIOS by disabling
technologies like speed stepping. However, it has to be ensured that
the processor has sufficient cooling under full load, which often is
not the case for mobile or Laptop devices. Alternatively, for Intel
CPUs, the older ACPI driver can be activated with boot parameter
\texttt{intel\_pstate=disable}, which supports to fix the frequency of
the processor.

\subsubsection{Controlling TLB Flushes and Shootdowns} 
Even though we have isolated cores from the scheduler, shootdowns may
still happen because the Linux kernel runs on the same core as the
process it is serving. Thus, every core that is used executes a part
of the kernel at some point, where shared kernel data can lead to
shootdowns.

In our kernel version KPTI is enabled, yet PCID support is not
available. Therefore, each mode switch flushes the TLB, significantly
skewing our measurements (see Section~\ref{sec:mode-switch}). We thus
disabled KPTI with kernel boot option \texttt{pti=off}\footnote{Note
  that this may open a security vulnerability!}.  This step is not
recommended when PCIDs are supported by both OS and CPU.

\subsubsection{Controlling Page Faults}
As earlier, we assume the user wants to avoid major page faults as far
as possible, since they have a penalty large enough to hide other
effects that may be of interest. Also, there is not much the user can
do if access to secondary storage is logically required, unless the
the program under analysis is redesigned.

In Linux, reading or writing data files is by default buffered through
the virtual memory. A read-ahead heuristic is used to bring data from
the slow disk to RAM, as soon as a demand is foreseen. However, while
this mechanism prevents major page faults quite effectively, it causes
DMA transfers and might not be desirable. Once a file is buffered in
virtual memory (``page cache''), data can be accessed without disk
access taking place. Obviously, the page cache can only prevent all
disk access if it is large enough to hold all files that will be in
use, and not flushed in between.

The easiest way to make use of the page cache is to run the
measurement twice, and discarding the results from the first
run. Alternatively, there exist tools which allow checking and
manipulating the page cache, e.g., \texttt{vmtouch}.


\subsubsection{Controlling Hardware Data Prefetching}
Although it is possible to distinguish some events by their cause
(i.e., demand vs. speculation), it cannot be known how much of the
prefetcher-induced penalties are hidden in out-of-order processing, or
how many of certain events are caused by the software directly.  One
shortcut to answer this question is to turn off hardware prefetching
and run the benchmarks twice, then compare results. This is specific
to every processor family, and may not be possible on all CPUs. For
Intel CPUs, machine-specific registers (MSRs) allow configuring the
processor in many ways, inter alia to turn off prefetchers.

\subsubsection{Controlling Influences of Peripherals}
If peripherals are sharing memory with the software under analysis, it
might be helpful to turn them off or minimize their impact, in case an
influence on the results cannot be ruled out. For example, the
graphics engine in Sandy Bridge can be made to relinquish parts of the
L3 cache by booting into a low-resolution mode.

\subsection{Taking Measurements}
The act of taking measurements itself is straightforward, and should
provide meaningful results if the previous recommendations have been
followed. We therefore summarize this only briefly.

\subsubsection{Performance Monitoring Units\label{sec:perf-monit-units}}
%
%
While there exist different ways and tools to measure performance,
such as hardware tracing and countless tools, we briefly describe the
use of the CPU's \emph{performance monitoring unit}
(PMU)~\cite{intelopt}, which provides hardware event counters.
These counters are processor-specific registers that are incremented
on the occurrence of certain events, for example, the number of L1
cache misses, or the number of processor cycles.  Under Linux, the
\texttt{perf} tool allows setting up and reading these registers, and
to separate event counts by process. A growing number of CPUs and SoCs offer
an equivalent to the PMU, e.g., many ARM Cortex processors already do.

Additional to hardware events, the \texttt{perf} also reads kernel
(software) counters, such as minor and major page faults, context
switches and CPU migrations.  \emph{Note:} We highly recommend to use
the native names of the registers as opposed to perf's names, to avoid
misunderstandings (e.g., the \texttt{perf} tool counts \emph{STLB
  hits} under the name \emph{ITLB loads} on Sandy Bridge), and to
watch out for errata (such as counter problems under HyperThreading).


\noindent\textbf{Multiplex and Grouping:} PMUs have a limited number of
hardware counters (eight, in our case). If we request more events than
counters, \texttt{perf} starts time-multiplexing, and extrapolating
the results from the sampling window to the entire life time of the
process. It is thus possible to miss certain events if the specific
counter is not currently active. Therefore, if the process behavior is
not stationary for long enough, the results may appear inconsistent.
One way around this issue is grouping of counters, which tries to
multiplex all counters of a group at the same time. However, some
processors have scheduling restrictions for which counters must not be
used together. One way around this is to avoid multiplexing at all,
and execute the program multiple times instead. This only makes
sense if the workload is repeatable. With Boolean options for
multiplexing (M) and grouping (G), there are four different
possibilities to measure, each with its own consequences:


\begin{enumerate}
\item \emph{Grouping and Multiplexing}: Possibly unsafe and incomplete
  results, because the process might be undersampled from M., and
  G. may create scheduling conflicts, which disables some counters.
\item \emph{No Grouping and Multiplexing}: Possibly unsafe but
  complete, because the process might be undersampled and counts
  inconsistent towards each other, but groups can be built in
  arbitrary ways to resolve scheduling conflicts.
\item \emph{No Multiplexing and no Grouping}: Possibly unsafe but
  complete, only works for repeatable workloads. Multiplexing,
  although not enabled, might still take place implicitly in an
  attempt to resolve conflicting counters.
\item \emph{No Multiplexing and Grouping}: Safe but possibly
  incomplete, only works for repeatable workloads. The grouping
  prevents implicit multiplexing from taking place.
\end{enumerate}

\noindent By \emph{unsafe} we mean that the counts may be both imprecise and
contradict each other (e.g., it might be possible that the L2 miss
count is greater than the L3 access count). By \emph{complete} we
mean that every event that has been requested was indeed counted at
some point in time.

Recently, \emph{weak} groups have been introduced in
\texttt{perf}~\cite{perfweak}, which can be broken to resolve
scheduling conflicts, but otherwise ensure that certain events are
counted synchronously.  It is thus a mixture of cases 1 and 2,
resulting in complete results with minimal multiplexing artifacts,
applicable to non-repeatable workloads.

We suggest to use grouping and avoid multiplexing as long as the
workload is repeatable, to obtain self-consistent and precise
results. Otherwise, grouping and multiplexing should be used,
preferably using weak groups.

For the system described in Tab.~\ref{tab:host}, the CPU supports
around 475 different events, offering deep insights into the
processor. However, looking at single counter values can be
misleading. Counters can have several orders of magnitude in
difference (e.g., page faults vs. cache misses), yet have a similar
impact on the resulting performance.  Especially when comparing two
versions of the same program, large differences may become meaningless
when compared to the absolute values.

\subsubsection{Hierarchical Bottleneck Analysis}
\label{Bottleneck Analysis}
\begin{figure}[htb]
  \centering
  \includegraphics[width=\columnwidth]{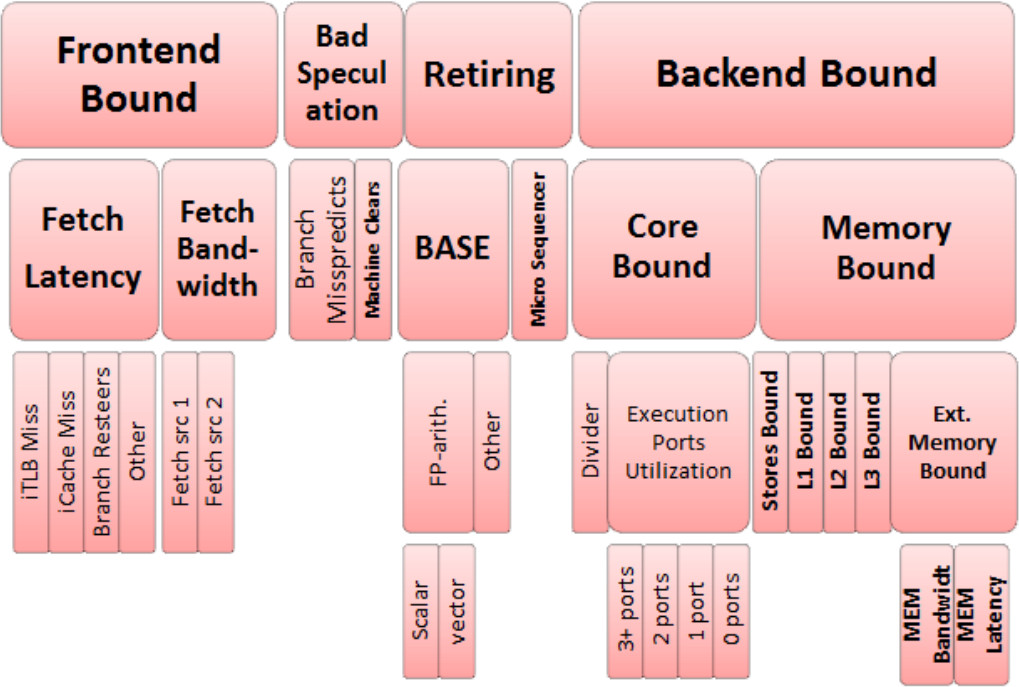}
  \caption{Hierarchical View on Processor
    Performance~\cite{DBLP:conf/ispass/Yasin14}.}
  \label{fig:yasin}
\end{figure}

To identify the true bottleneck of an application, Yasin has proposed
a hierarchical analysis~\cite{DBLP:conf/ispass/Yasin14}. The analysis
follows the hierarchical anatomy of general OoO processors, depicted
in Fig.\ref{fig:yasin}. As a first level, the user should only be
concerned whether the majority of cycles is spent in the Frontend
(entails decoding instructions, L1d access and Microcode switches), or
in Bad Speculation (machine clears and branch mispredictions) or in
the Backend (L1/2/3 and DRAM data access) or whether most of the time
is spent in retiring instructions (ideally, 100\%). Once the most
time-consuming category is identified, the user should focus on its
children at the next level, determine the most prominent one, and so
on. For a full explanation, the reader is referred
to~\cite{DBLP:conf/ispass/Yasin14}.

A free implementation of this analysis for Intel CPUs is available in
a tool called \texttt{toplev}, which is part of Intel's
\texttt{pmu-tools}~\cite{pmutools}. It takes measurements using Linux'
\texttt{perf} tool, and presents the results in the explained
hierarchy, but in ratios rather than absolute numbers. Multiplexing
and grouping is supported as described earlier. Taking the
measurements thus boils down to invoking the tool. 

Only after the bottleneck has been identified, individual counters
should be inspected using \texttt{perf}, since now we know they are
meaningful.  Last but not least, it is worth noting that
\texttt{pmu-tools} also includes a convenience wrapper
for \texttt{perf}, called \texttt{ocperf}. This tool allows using
native names for Intel CPUs, and has better support for uncore events.

It should be noted that \texttt{toplev} does not allow localizing
causes of undesired behavior in the source code, since counters
operate cumulatively and are not associated to specific locations in
the source code of a program. A method for localization is provided by
\texttt{perf} in \emph{sampling} mode, where a history of events can
be logged, annotated in the source code. The tool 
offers many options explained in the documentation~\cite{perfwiki}.

\subsubsection{Uncertainty Propagation}
All measured events can be associated with a measurement
uncertainty. This uncertainty indicates the precision of the measured
values, and prevents us from drawing false conclusions if we are
comparing two versions of a software.

Crucially, not all events can be measured directly on the hardware,
and thus must be calculated from others, measured or themselves calculated,
ones. The calculated events therefore have an uncertainty based on
their constituents, which needs to be properly tracked. 
As an example, on Sandy Bridge the cache hit ratio $r$ must be
estimated from access $a$ and hit $h$ counters, and its standard
deviation $\sigma_r$ depends on both measurements as follows
\begin{equation}
  \sigma_r = \left|\frac{h}{a}\right|\sqrt{\left(\frac{\sigma_h}{h}\right)^2 + \left(\frac{\sigma_a}{a}\right)^2 - 2\frac{\sigma_{a,h}}{ah}},
\end{equation}
where $\sigma_{a,h}$ is the covariance between accesses and
hits. Analogously, the uncertainty is propagated for multiplication,
division, addition, and all other operations~\cite{ku1966notes}.

We have contributed patches to \texttt{toplev}, which track the
standard deviation of counters across multiple runs
(\texttt{--repeat)} through all calculations, while assuming
statistically independent variables, i.e., $\sigma_{a,h}=0$.  The
resulting uncertainty for each event is indicated with error bars in
our plots.

\emph{Warning:} When measurements are taken in system-wide mode across
more than one logical processor core, then \texttt{toplev} currently
forces \texttt{perf} to not aggregate the results (flag
\texttt{-A}). This in turn suppresses \texttt{perf}'s output of the
standard deviation, and leads to an optimistic
uncertainty. System-wide mode is also forced when HyperThreading (HT)
is activated, thus HT suppresses the standard deviation, as
well. There are thus three options to capture measurement uncertainty
correctly:
\begin{inparaenum}[(1)]
\item \textbf{Recommended:} Disable HT and avoid system-wide mode. The
  \texttt{perf} tool will ``follow'' the software under analysis to
  whatever core it is going (if not pinned).
\item \textbf{Alternative 1:} Disable HT and use system-wide mode
  while filtering only for only one single core. The software under
  analysis must be pinned to that single core.
\item \textbf{Alternative 2:} Keep HT and use system-wide mode, and
  specify \texttt{--single-thread}, which forces \texttt{toplev} to
  aggregate the results across all CPUs. The system should otherwise
  be idle.
\end{inparaenum}
In all cases, the core where the software is running on should be
isolated, as explained in Section~\ref{sec:experiments}.  Future
releases of \texttt{perf}/\texttt{toplev} might no longer have this
caveat.

\section{Experiments}
%
%
We provide a few short examples illustrating the difference between
our proposed measurement setup, and a default Linux environment. 
We have chosen three programs with different characteristics:
\begin{itemize}
\item \emph{gnugo}: This is a game engine for the Chinese Go game. The
  program consists of many branches and jumps, many of which are hard
  to predict. Additionally, it has a low spatial locality of
  instructions, and will therefore like suffer from many instruction
  cache misses. Memory usage is low compared to the next program.
\item \emph{stream}: This is McCalpin's memory bandwidth
  benchmark~\cite{mccalpin1995sustainable}, which stresses the memory
  subsystem heavily, but in a predictable pattern. Consequently, this
  program occupies a lot of memory (1.1GB in our case), but has few
  branches or jumps.
\item \emph{syscalls}: This is the program shown in
  Appendix~\ref{sec:mode-switch-benchm}, which exercises syscalls. 
  The memory usage is low compared to the others, but it
  causes many mode switches and thus allows more context switches to take
  place. This program should keep the functional units busy.
\end{itemize}

All programs have been executed five times in a row to allow
determining a standard deviation, which is then used for our uncertainty
propagation. We have repeated this for two different measurement
setups; first with the default settings of the system described in
Tab.~\ref{tab:host}, and another time with our proposed measurement
setup described in Section~\ref{sec:experiments}.

All measurements have been taken with similar parameters, using the
performance monitoring counters (see
Section~\ref{sec:perf-monit-units}. Specifically, multiplexing and
HyperThreading were disabled, to prevent undersampling and to ensure
proper uncertainty propagation, as explained in
Section~\ref{Bottleneck Analysis}.

\begin{figure}[htb]
  \centering
  \input{img/fig_classes}
  \caption{Results of hierarchical performance analysis for default
    setup (dfl) and our proposed setup (our).}
  \label{fig:classes}
\end{figure}
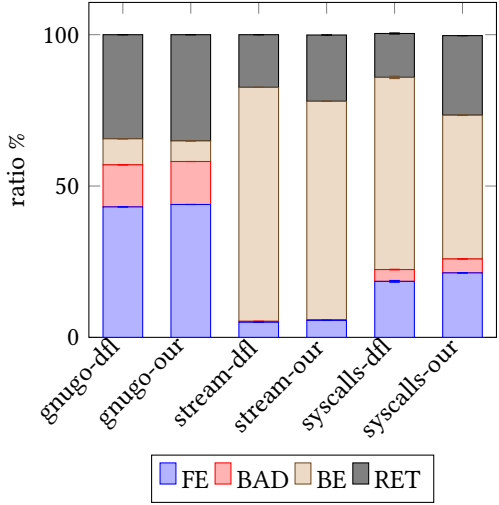

\subsection{Hierarchical Bottleneck Analysis}
Figure~\ref{fig:classes} shows the results of the first level of the hierarchical
analysis using \texttt{toplev}.  It can be seen that the measurements
all programs reflect the expected behavior: \emph{gnugo} indeed spends
a lot of time in its frontend (FE), due to many cache misses and
branch mispredictions (see Fig.\ref{fig:gnugodetail}). The
\emph{stream} benchmark spends most of its time in the backend (BE),
because it is mostly memory-bound (see
Fig.\ref{fig:streamdetail}). The \emph{syscalls} program shows itself
to be mainly backend-heavy, but this time due to core usage (see
Fig.\ref{fig:syscallsdetail}).

The difference between the default setup and ours is however barely
visible for this type of analysis. Only \emph{syscalls} shows a larger
difference. The results with our proposed setup make the program
appear less back-end bound, in exchange for more time spent retiring
instructions. This hierarchical analysis does not offer any more
explanation, and thus we will revisit this in the next section.

All figures have error bars, but only very few of them are visible due
to their low magnitude. This suggests that a default setup may be
sufficient in terms of measurement uncertainty, and that the results
of a hierarchical analysis may be close enough between the two setups,
at least for the programs tested here. This is surprising, because the
absolute values of the counters, as well as program performance,
differ significantly, as we show next.

\begin{figure*}[htb]
  \centering
\begin{lstlisting}
FE        Frontend_Bound:                                    43.94 +- 0.02 % Slots             <==
BAD       Bad_Speculation:                                   14.16 +- 0.01 % Slots                
RET       Retiring:                                          35.07 +- 0.01 % Slots below          
FE        Frontend_Bound.Frontend_Latency:                   22.77 +- 0.03 % Slots                
FE        Frontend_Bound.Frontend_Bandwidth:                 21.13 +- 0.03 % Slots                
BAD       Bad_Speculation.Branch_Mispredicts:                14.07 +- 0.01 % Slots                
FE        Frontend_Bound.Frontend_Latency.ICache_Misses:     17.66 +- 0.00 % Clocks_Estimated     
FE        Frontend_Bound.Frontend_Latency.Branch_Resteers:   14.07 +- 0.01 % Clocks_Estimated     
\end{lstlisting}  
  \caption{\texttt{toplev} output for \emph{gnugo} with our proposed
    measurement setup.\label{fig:gnugodetail}}
\end{figure*}

\begin{figure*}[htb]
  \centering
\begin{lstlisting}
FE        Frontend_Bound:                                     5.62 +- 0.00 % Slots below           
BAD       Bad_Speculation:                                    0.24 +- 0.00 % Slots below           
BE        Backend_Bound:                                     72.20 +- 0.00 % Slots                 
RET       Retiring:                                          21.94 +- 0.00 % Slots below           
BE/Mem    Backend_Bound.Memory_Bound:                        52.55 +- 0.02 % Slots             <==
BE/Core   Backend_Bound.Core_Bound:                          19.65 +- 0.02 % Slots                 
BE/Mem    Backend_Bound.Memory_Bound.L1_Bound:                0.00 +- 0.00 % Stalls below          
BE/Mem    Backend_Bound.Memory_Bound.L2_Bound:                1.73 +- 0.02 % Stalls below          
BE/Mem    Backend_Bound.Memory_Bound.L3_Bound:               17.80 +- 0.03 % Stalls below          
BE/Mem    Backend_Bound.Memory_Bound.DRAM_Bound:             36.13 +- 0.04 % Stalls below          
BE/Mem    Backend_Bound.Memory_Bound.Store_Bound:            10.27 +- 0.01 % Stalls below          
BE/Core   Backend_Bound.Core_Bound.Ports_Utilization:        21.92 +- 0.03 % Clocks                
\end{lstlisting}  
  \caption{\texttt{toplev} output for \emph{stream} with our proposed
    measurement setup.\label{fig:streamdetail}}
\end{figure*}

\begin{figure*}[htb]
  \centering
\begin{lstlisting}
FE        Frontend_Bound:                                    21.40 +- 0.05 % Slots                 
BE        Backend_Bound:                                     47.52 +- 0.08 % Slots                 
RET       Retiring:                                          26.31 +- 0.01 % Slots below           
FE        Frontend_Bound.Frontend_Latency:                   12.94 +- 0.03 % Slots below           
BE/Mem    Backend_Bound.Memory_Bound:                         0.00 +- 0.00 % Slots below           
BE/Core   Backend_Bound.Core_Bound:                          47.52 +- 0.08 % Slots                 
BE/Core   Backend_Bound.Core_Bound.Ports_Utilization:        39.88 +- 0.20 % Clocks            <==
\end{lstlisting}  
  \caption{\texttt{toplev} output for \emph{syscalls} with our proposed
    measurement setup.\label{fig:syscallsdetail}}
\end{figure*}

\subsection{Absolute Event Counts and Absolute Performance}
Hierarchical analysis has not been showing any larger differences
between the measurement setups. It builds on ratios rather than
absolute values, therefore large differences could become invisible.
While this is acceptable and desirable for bottleneck analysis, it
might be undesirable in other scenarios, such as when absolute
performance or the precise event counts are required.  These absolute
counts differ significantly, as we show in the following.

\begin{figure*}[htb]
  \centering
  \includegraphics[width=.93\textwidth]{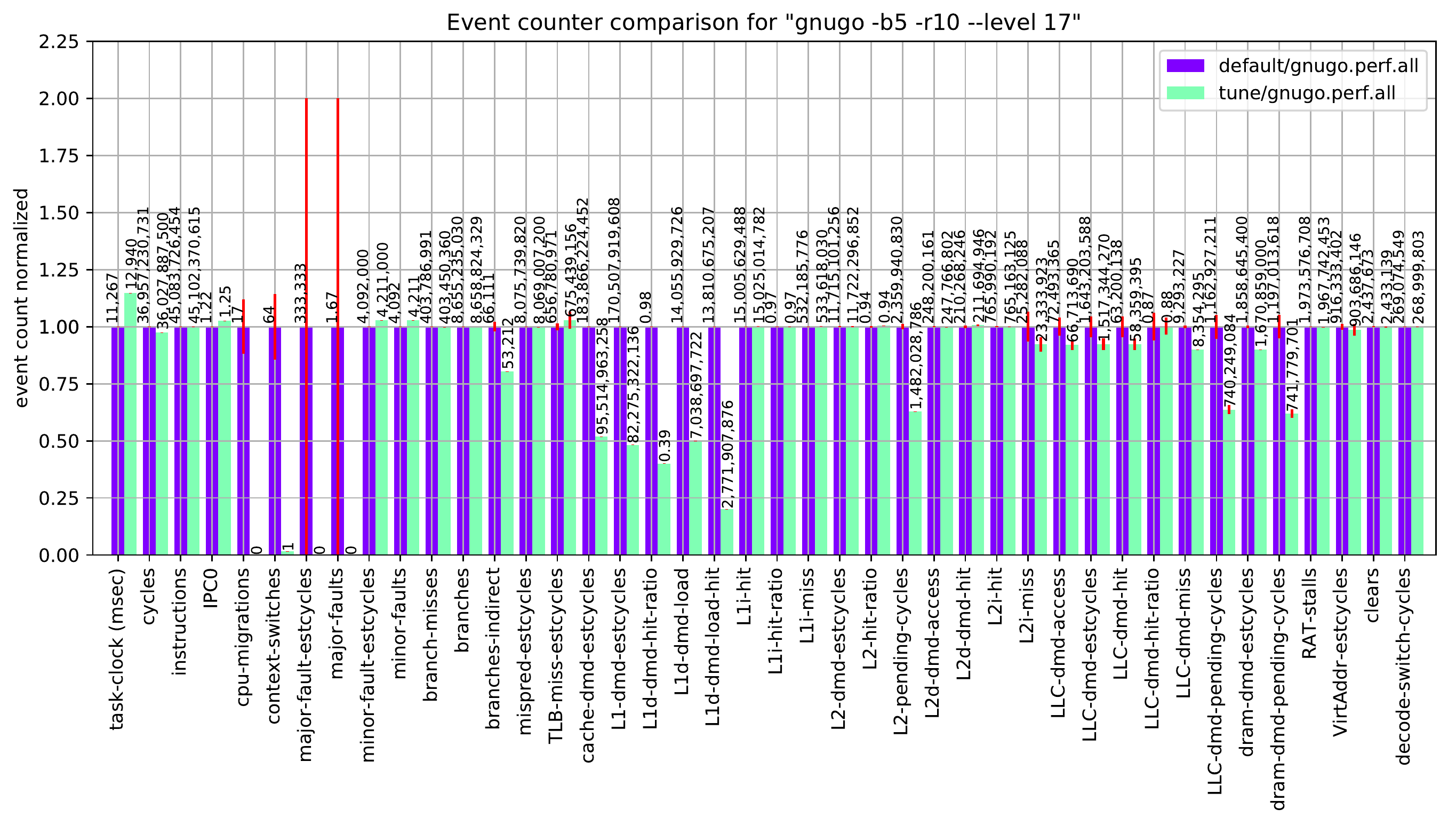}
  \caption{Absolute event counts for \emph{gnugo} for default setup
    (default) and our proposed setup (tune). Red error bars indicate
    measurement uncertainty.}
  \label{fig:abs_gnugo}
\end{figure*}

Figure~\ref{fig:abs_gnugo} shows the absolute event counts of
\emph{gnugo} for our proposed setup (``tune'') and for the default
one. First, it can be seen that the program runs faster under the
default setup. The bar ``task-clock (msec)'' shows 11.4s vs 12.9s;
note that task-clock only increments for those cycles where the
program has been active on the processor, which is always less or
equal than wall-clock time. As a better metric reflecting wall-clock
time, all plots show a bar called ``IPC0'', which is calculated as
$I/(f\cdot t_w)$, where $I$ is the number of instructions, $f$ the
processor frequency, and $t_w$ the wall-clock time of the program. In
Fig.\ref{fig:abs_gnugo}, IPC0 is almost identical for both measurement
setups, thus the programs would take about the same time to execute in
both setups.

The next apparent difference is the absence of context switches (one
switch is always necessary) and CPU migrations in our setup, followed
by no major faults in our setup, and a different L1d caching
behavior. All in all, the measurement uncertainty again is comparable
when the counts are close.

The results for \emph{stream} and \emph{syscalls} are shown in
Appendix~\ref{sec:absol-event-counts}. Again, there are some
differences in the absolute event counts. This time, IPC0 shows that
\emph{stream} runs faster in our setup (~$0.6/0.8=25\%$), whereas
\emph{syscalls} is significantly slower (~$0.68/0.15=450\%$), as
expected due to the higher overhead for mode switches. Furthermore,
\emph{syscalls} also shows large differences in most counters. The
entire characteristics seem to be driven by the differences in the
mode switch cost, and even outweighs that the default setup performs
many more context switches and CPU migrations than our proposed setup.

In conclusion, the measurement setup can make a large difference in
absolute event counts and software performance. Our experiments
suggest that neither setup is always dominating in performance, and
that the choice of setup should depend on the software under analysis.

\subsection{Stability towards Background Processes}
So far, all measurements have been taken on an idle system, with no
interactive user processes being executed. We now examine how
measurements change when we run some background processes, as often
the case in production systems. Specifically, we run four background
processes which stress caches and DRAM (\texttt{stress-ng -C2 --vm 2
 --vm-populate --vm-bytes=512m}).
\label{sec:stab}

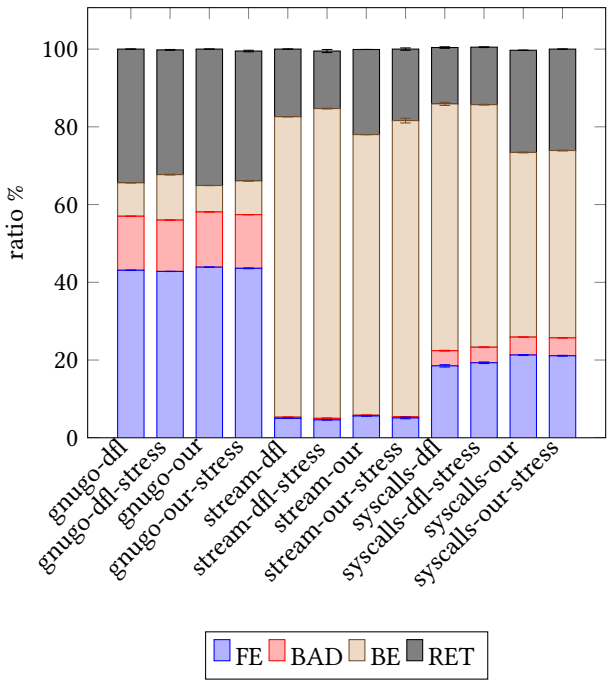
\begin{figure}[htb]
  \centering
  \input{img/fig_robust}
  \caption{Results of hierarchical performance analysis with and
    without highly active background processes, for default setup (dfl)
    and our proposed setup (our).}
  \label{fig:robust}
\end{figure}

\subsubsection{Hierarchical Analysis}
Figure~\ref{fig:robust} shows the results of a hierarchical analysis,
which certify that all measurements under all setups are stable. There
are small differences, which however do not change the results
fundamentally. Again, the measurement uncertainty surprisingly is
always very low. The results suggest that the hierarchical analysis
is also robust against background processes, at least at the uppermost
level of the hierarchy.

\subsubsection{Absolute Event Counts and Absolute Performance}
The absolute event counts give a similar picture as before. However,
the wall-clock time of the programs is considerably more stable in our
setup.

\begin{figure*}[htb]
  \centering
  \includegraphics[width=.93\textwidth]{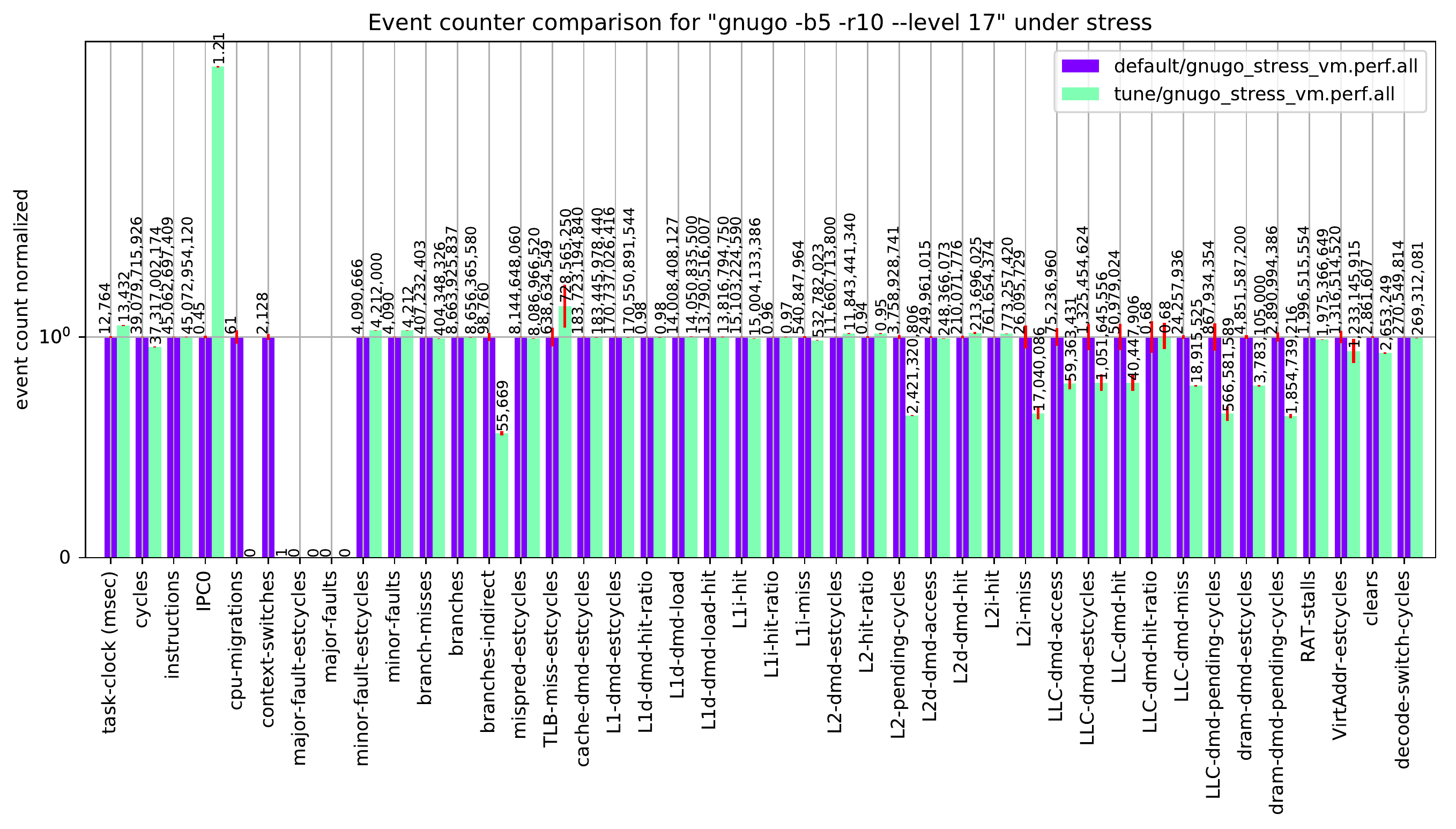}
  \caption{Absolute event counts for \emph{gnugo} for default setup
    (default) and our proposed setup (tune) with background processes. Red
    error bars indicate measurement uncertainty. Large performance
    differences are visible.}
  \label{fig:absperf_go}
\end{figure*}

Figure~\ref{fig:absperf_go} shows again absolute event counts for both
measurement setups of \emph{gnugo}, when background processes are
running. A large change in IPC0 can be seen. In our setup, the
program almost maintains the same performance as on an idle system
(4\% degradation, since the IPC0 drops from 1.25 to 1.21), whereas in
the default setup performance significantly degrades (64\%
degradation, since IPC0 drops from 1.22 to 0.45).  The result is
similar for \emph{stream}, which degrades by 16\% in our setup, but by
67\% in the default setup, and for \emph{syscalls}, which degrades by
60\% in the default setup, but only by 2\% in our setup.

In summary, the absolute event counts are considerably different in
our setup, and additionally the performance of the program under
background load is significantly better maintained in our setup. 

\subsection{Other Influences}
When performance counters are used, there is a low but visible
overhead, as with many other measurement methods. We have neglected
this overhead in our experiments.

\section{Related Work}

\noindent\textbf{Hardware Performance Counters:} Since all performance counters
are vendor- and processor-specific, different profiling software
exists to access them. A generic tool is Linux' \texttt{perf}, which
also supports some many AMD, ARM, ARC, Blackfin, MIPS, PowerPC and
SPARC processors~\cite{perfarch}. Specific to Intel CPUs, we have
\texttt{toplev}, which we have mentioned earlier~\cite{pmutools}, and
the commercial tool \emph{VTune}. Interestingly, some processors also
provide counters for power
events~\cite{Singh:2009:RTP:1577129.1577137}, for which our setup
might be equally important.
\noindent\textbf{Software-based Measurements:} There exist many
well-known tools for debugging software performance. Perhaps best
known are software profilers, such as \texttt{gprof} (execution time)
and Visual Studio Profiler. More generally, there exists the larger
class of dynamic program analyzers, which observe programs during
run-time, sometimes by instrumentation, other times through
simulation.  A survey about those was recently given in
\cite{gosain2015survey}. One successful framework worth mentioning is
Valgrind~\cite{nethercote2007valgrind}, which provides tools to track
execution time and memory usage. However, such tools change the
program's characteristics through instrumentation or simulation, and
furthermore do not provide any microarchitectural
explanations. Finally, we should also mention \texttt{ftrace}, which
can collect software events during the execution of a program, and is
integrated with \texttt{perf}.

\noindent\textbf{Reducing OS Noise}: Another attempt at reducing
the OS' impact on workloads has been described by Akkan
et. al~\cite{akkan2013understanding}.
%
However, their report is mainly concerned with the OS, and furthermore
neglects some important effects (e.g., HyperThreading, \texttt{ftrace} overhead
and tick length). Nevertheless, some pointers can be found there.  For
technical details in various aspects on Linux, the \emph{Linux Kernel
  Mailing List} (LKML), the kernel docs, and LWN have been invaluable
primary sources for understanding the inner workings of the kernel. We
have given the references where applicable, and many more can be found
by the interested reader. Last but not least, to erase even the
faintest doubt of imprecise or deprecated documentation, the Linux
source code itself serves as a definitive reference.



\section{Concluding Remarks}
We have described the most important features of both the Linux
operating system and a modern out-of-order superscalar processor, and
how they influence software performance. Based on the identified
influences, we have proposed a measurement setup that aims to reduce
measurement errors, such that not the OS or hardware is characterized,
but mainly the program under analysis. Towards that, we have extended
a hierarchical performance analysis method with uncertainty
propagation, to indicate the measurement errors.

Surprisingly, our
experiments showed that for this hierarchical and ratio-based
performance analysis method, the measurement setup makes little
difference. In contrast, when absolute event counts are of interest,
e.g., when cache access counts shall be correlated to source code, our
setup has shown significant improvements.

As a side effect, we have found that the proposed setup is more immune to
background processes than a default setup. Programs show considerably
better stability in their observable performance, whereas a
performance degradation of up to 64\% has been observed in a default
setup. Our proposed setup can therefore be used as a guideline to tune
the system for high-performance applications.

Naturally, the results presented here may change with different
processors and OS versions. We have given references for the reader to evaluate which
parts of the setup may become relevant, but an extrapolation of the
presented results to all targets would be unsafe.

In conclusion, the ultimate measurement setup does not exist. It
depends on both the requirements for the measurements and the
characteristics of the software under analysis. For more robust
measurements, we recommend our proposed setup. This may however
degrade performance for some programs, and thus a subset of the
presented configurations might be chosen to achieve optimal results.

\appendix
{
 \bibliographystyle{ACM-Reference-Format}
 \bibliography{literature}
}

\pagebreak
\section{Mode Switch Benchmark for x86\label{sec:mode-switch-benchm}}
\begin{lstlisting}
#include <x86intrin.h>
#include <stdio.h>
#include <unistd.h>

#define TWENTY_MILLION 20000000
#define NTRIALS TWENTY_MILLION

int main(void) {
    unsigned long ini, end, now, best, tsc;
    int i;

#define measure_time(code) \
    for (i = 0; i < NTRIALS; i++) { \
        ini = __rdtsc(); \
        code; \
        end = __rdtsc(); \
        now = end - ini; \
        if (now < best) best = now; \
    }

    /* time rdtsc itself (i.e. no code) */
    best = ~0;
    measure_time (0);
    tsc = best;
    printf ("rdtsc: %li cycles\n", tsc);
    
    /* time one of the fastest syscalls */
    best = ~0;
    measure_time (getuid());
    printf ("getuid(): %li cycles\n", best-tsc);
    return 0;
}
\end{lstlisting}

\section{Absolute Event Counts\label{sec:absol-event-counts}}
\begin{figure*}[htbp]
  \centering
  \includegraphics[width=.9\textwidth]{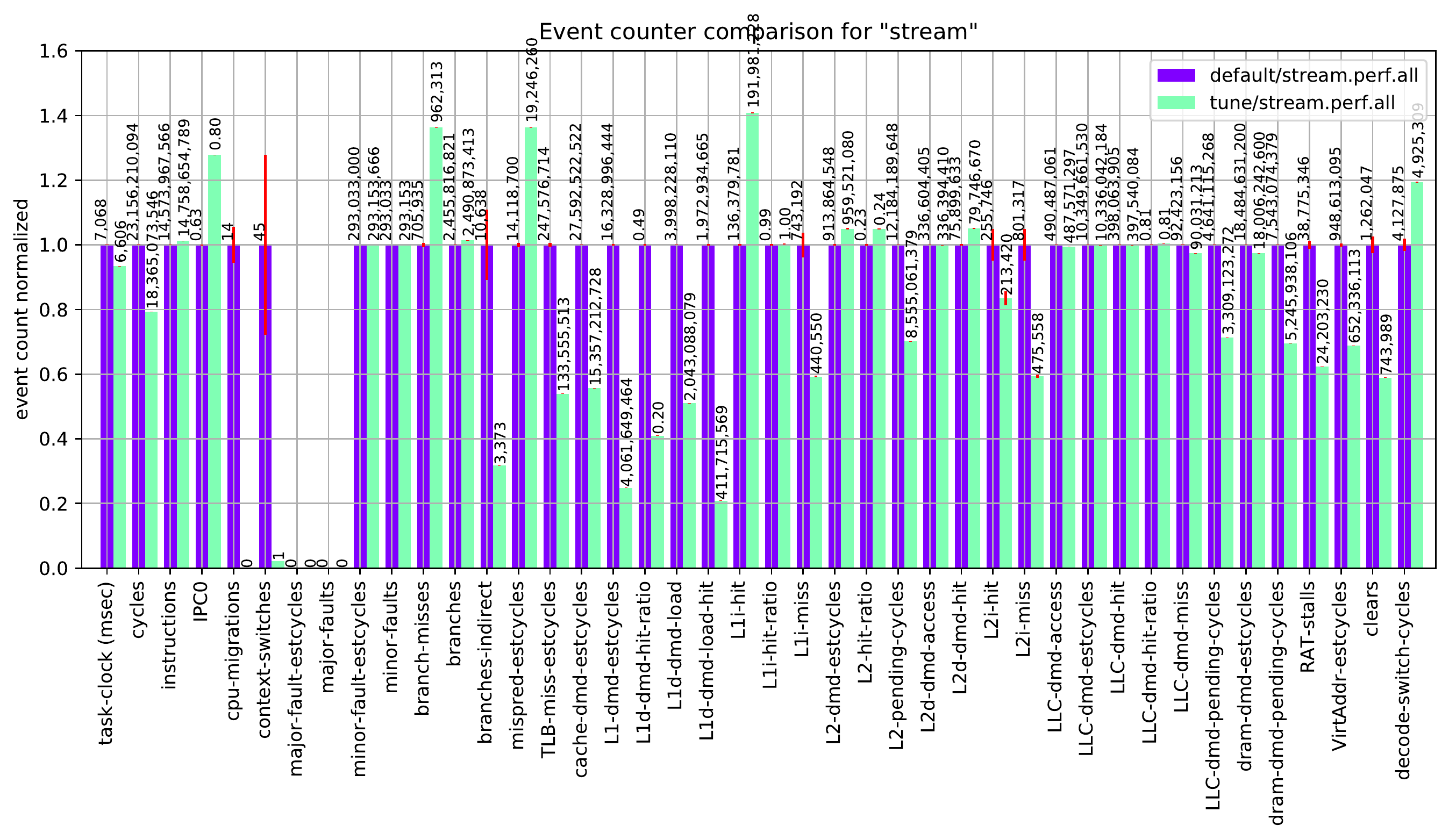}
  \caption{Absolute event counts for \emph{stream} for default setup
    (default) and our proposed setup (tune).}
  \label{fig:abs_stream}
\end{figure*}

\begin{figure*}[htbp]
  \centering
  \includegraphics[width=.9\textwidth]{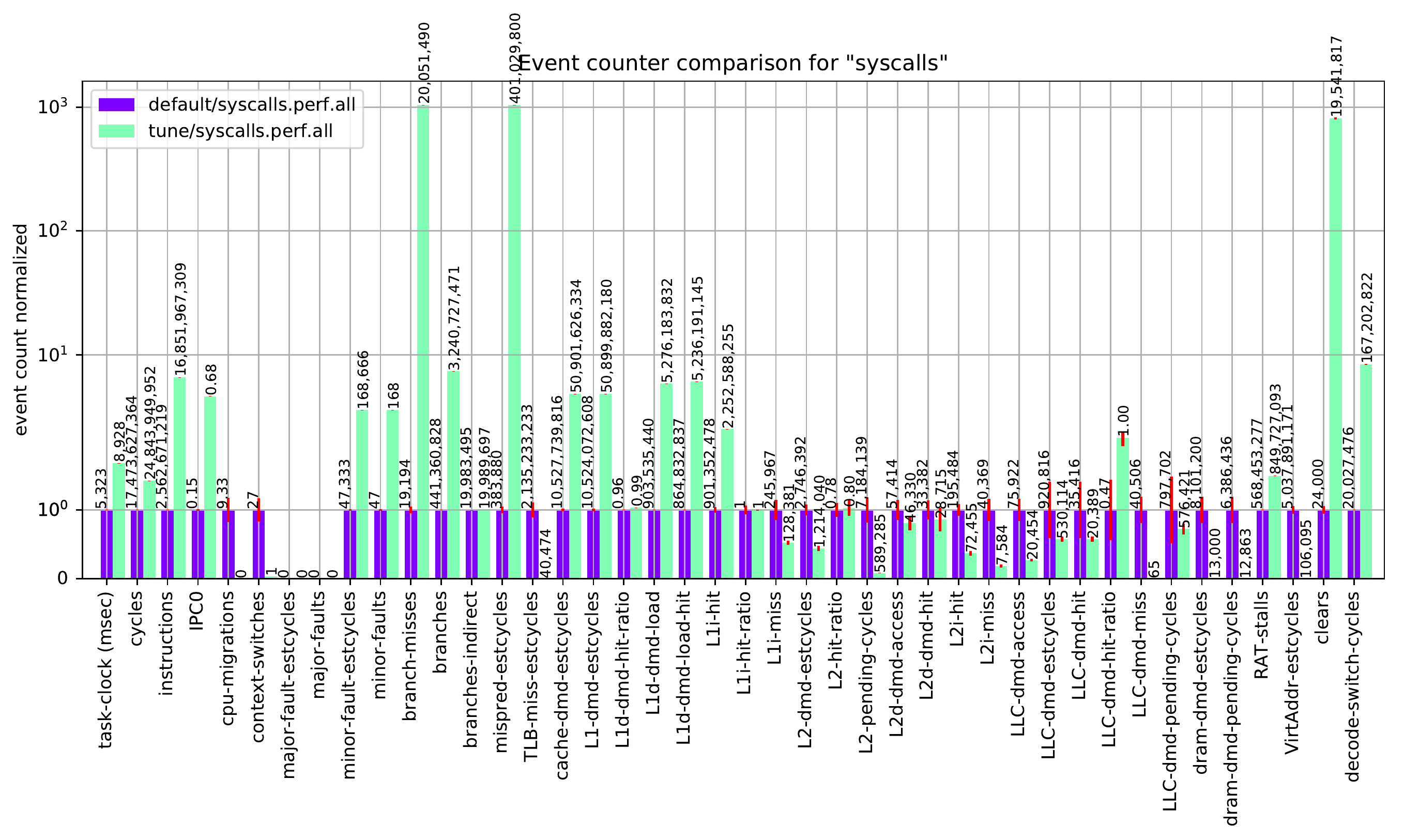}
  \caption{Absolute event counts for \emph{syscalls} for default setup
    (default) and our proposed setup (tune).}
  \label{fig:abs_syscalls}
\end{figure*}

\begin{figure*}[htbp]
  \centering
  \includegraphics[width=.9\textwidth]{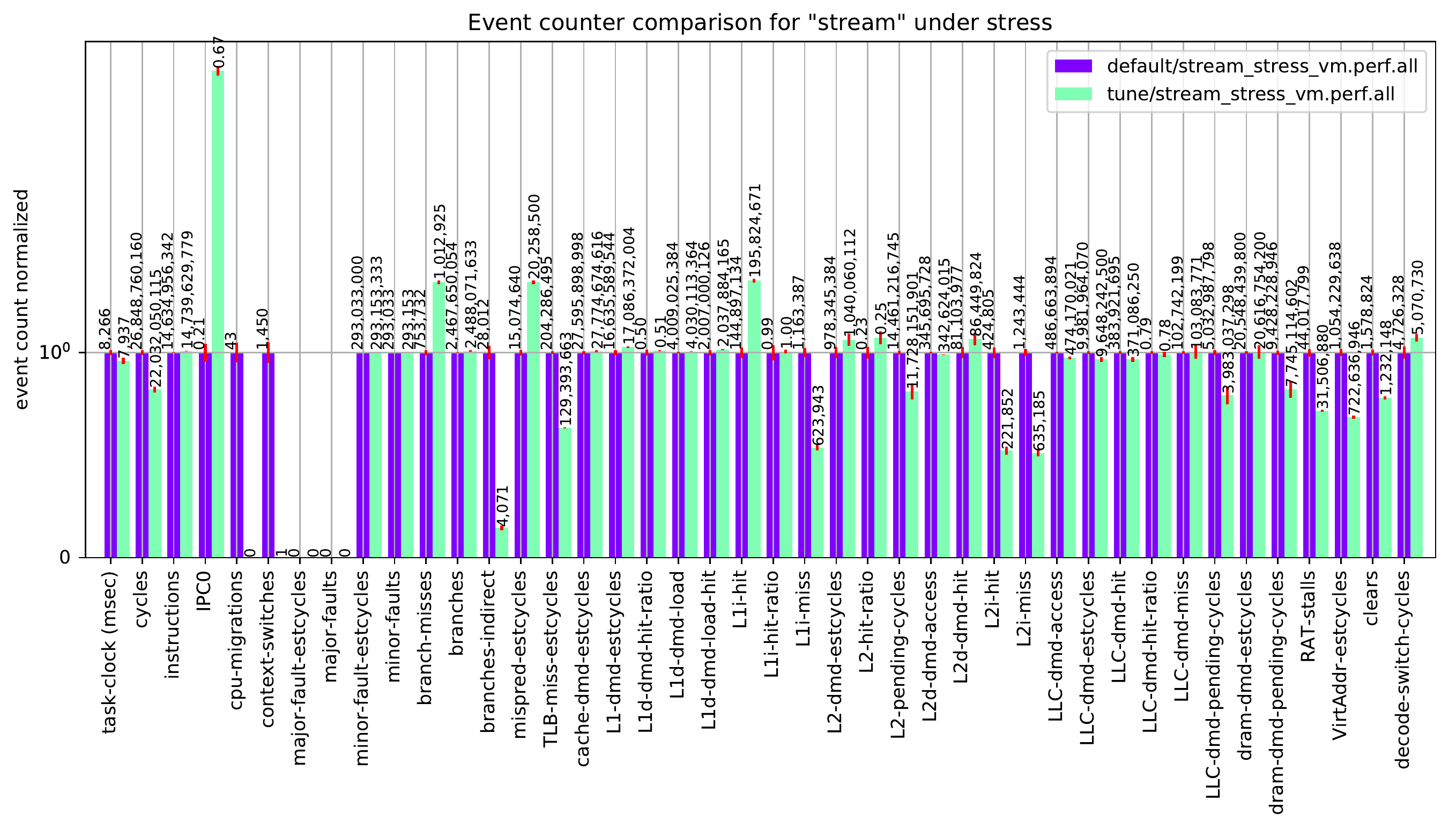}
  \caption{Absolute event counts for \emph{stream} for default setup
    (default) and our proposed setup (tune) with background processes.}
  \label{fig:absperf_stream}
\end{figure*}

\begin{figure*}[htbp]
  \centering
  \includegraphics[width=.9\textwidth]{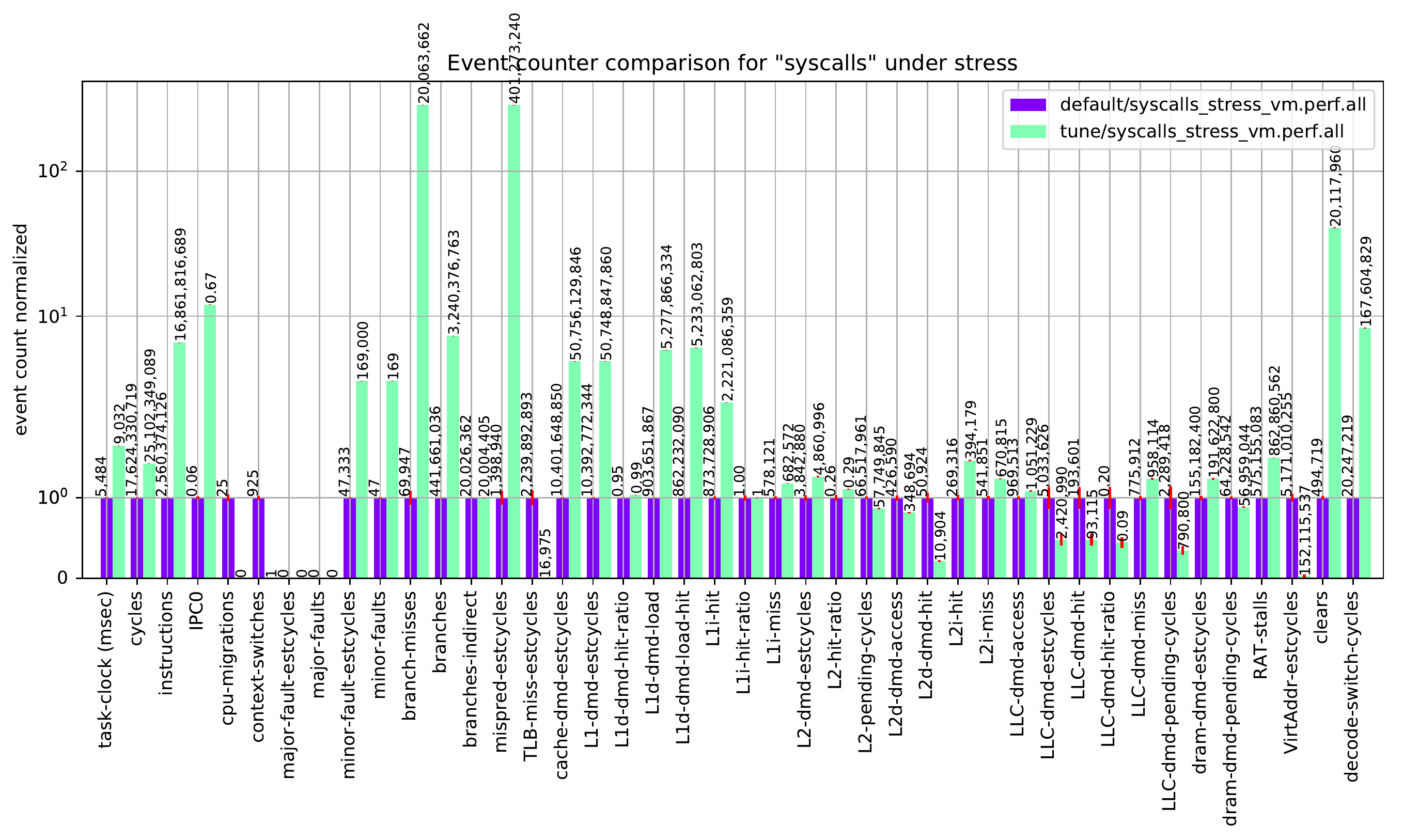}
  \caption{Absolute event counts for \emph{syscalls} for default setup
    (default) and our proposed setup (tune) with background processes.}
  \label{fig:absperf_syscalls}
\end{figure*}

\end{document}

%% file: tumci.tex
\usepackage{color}

\definecolor{TUMblue}{RGB}   {  0, 101, 189} 
\definecolor{Pantone540}{RGB}{  0,  51,  89}
\definecolor{Pantone301}{RGB}{  0,  82, 147}
\definecolor{Pantone285}{RGB}{  0, 115, 207}
\definecolor{Pantone542}{RGB}{100, 160, 200}
\definecolor{Pantone283}{RGB}{152, 198, 234}
\definecolor{TUMdarkgray}{RGB}{ 88, 88, 90}
\definecolor{TUMgray}{RGB}{ 156, 157, 159}
\definecolor{TUMlightgray}{RGB}{ 217, 218, 219}
\definecolor{TUMgreen}{RGB}{162, 173, 0} 
\definecolor{TUMorange}{RGB}{227, 114, 34} 
\definecolor{TUMelfenbein}{RGB}{218, 215, 203} 
\definecolor{TUMpyellow}{RGB}{255, 180, 0}
\definecolor{TUMporange}{RGB}{255, 128, 0}
\definecolor{TUMpred}{RGB}{229, 52, 24}
\definecolor{TUMpdarkred}{RGB}{202, 33, 63}
\definecolor{TUMpblue}{RGB}{0, 153, 255}
\definecolor{TUMplightblue}{RGB}{65, 190, 255}
\definecolor{TUMpgreen}{RGB}{145, 172, 107}
\definecolor{TUMplightgreen}{RGB}{181, 202, 130}
\definecolor{TUMpbluethemelower}{RGB}{0, 82, 147} 
\definecolor{TUMpbluethemeupper}{RGB}{0, 40, 72}


%% file: img/fig_classes.tex
  \begin{tikzpicture}
    \begin{axis}[
      ybar stacked,
      bar width=15pt,
      legend style={at={(0.5,-0.35)},
        anchor=north,legend columns=-1},
      ylabel={ratio \%},
      symbolic x coords={gnugo-dfl, gnugo-our, stream-dfl, stream-our, syscalls-dfl, syscalls-our},
      xtick=data,
      ymin=0,
      x tick label style={rotate=45,anchor=east},
      ]
      
      \addplot+[ybar,error bars/.cd,y dir=both,y explicit] plot 
      coordinates {
        (gnugo-dfl,43.1) +- (0,0.03) 
        (gnugo-our,43.9)  +- (0,0.02) 
        (stream-dfl,5.0) +- (0,0.02) 
        (stream-our,5.6) +- (0,0) 
        (syscalls-dfl,18.5) +- (0,0.3) 
        (syscalls-our,21.3) +- (0,0.1) 
      };

      \addplot+[ybar,error bars/.cd,y dir=both,y explicit] plot 
      coordinates {
        (gnugo-dfl,13.9) +- (0,0.02) 
        (gnugo-our,14.2)  +- (0,.01)
        (stream-dfl,0.3) +- (0,0.01)
        (stream-our,0.2) +- (0,0)
        (syscalls-dfl,3.9) +- (0,0.1)
        (syscalls-our,4.6) +- (0,0.0)
      };

      \addplot+[ybar,error bars/.cd,y dir=both,y explicit] plot 
      coordinates {
        (gnugo-dfl,8.6) +- (0,0.04)
        (gnugo-our,6.8) +- (0,0.03)
        (stream-dfl,77.3) +- (0,0.02)
        (stream-our,72.2) +- (0,0)
        (syscalls-dfl,63.5) +- (0,0.4)
        (syscalls-our,47.5) +- (0,0.1)
      }; 

      \addplot+[ybar,error bars/.cd,y dir=both,y explicit] plot 
      coordinates {
        (gnugo-dfl,34.4) +- (0,0)
        (gnugo-our,35.1) +- (0,0.01)
        (stream-dfl,17.4) +- (0,0)
        (stream-our,21.9) +- (0,0)
        (syscalls-dfl,14.5) +- (0,0.2)
        (syscalls-our,26.3) +- (0,0)
      }; 

      \legend{\strut FE, \strut BAD, \strut BE, \strut RET}
    \end{axis}
  \end{tikzpicture}

%% file: img/fig_robust.tex
  \begin{tikzpicture}
    \begin{axis}[
      ybar stacked,
      width=\columnwidth,
      bar width=10pt,
      legend style={at={(0.5,-0.45)},
        anchor=north,legend columns=-1},
      ylabel={ratio \%},
      symbolic x coords={gnugo-dfl, gnugo-dfl-stress, gnugo-our, gnugo-our-stress, stream-dfl, stream-dfl-stress, stream-our, stream-our-stress, syscalls-dfl, syscalls-dfl-stress, syscalls-our, syscalls-our-stress},
      xtick=data,
      ymin=0,
      x tick label style={rotate=45,anchor=east},
      ]
      
      \addplot+[ybar,error bars/.cd,y dir=both,y explicit] plot 
      coordinates {
        (gnugo-dfl,43.1) +- (0,0.03) 
        (gnugo-dfl-stress,42.8) +- (0,.03)  
        (gnugo-our,43.9)  +- (0,0.02)
        (gnugo-our-stress,43.6)  +- (0,0.1) 
        (stream-dfl,5.0) +- (0,0.02) 
        (stream-dfl-stress,4.6) +- (0,0.03) 
        (stream-our,5.6) +- (0,0)
        (stream-our-stress,5.1) +- (0,0.2) 
        (syscalls-dfl,18.5) +- (0,0.3)
        (syscalls-dfl-stress,19.3) +- (0,0.2) 
        (syscalls-our,21.3) +- (0,0.1)
        (syscalls-our-stress,21.1) +- (0,0.1) 
      };

      \addplot+[ybar,error bars/.cd,y dir=both,y explicit] plot 
      coordinates {
        (gnugo-dfl,13.9) +- (0,0.02) 
        (gnugo-dfl-stress,13.2) +- (0,0.03) 
        (gnugo-our,14.2)  +- (0,.01)
        (gnugo-our-stress,13.8)  +- (0,0) 
        (stream-dfl,0.3) +- (0,0.01)
        (stream-dfl-stress,0.4) +- (0,0.1) 
        (stream-our,0.2) +- (0,0)
        (stream-our-stress,0.3) +- (0,0) 
        (syscalls-dfl,3.9) +- (0,0.1)
        (syscalls-dfl-stress,4.0) +- (0,0.1) 
        (syscalls-our,4.6) +- (0,0.0)
        (syscalls-our-stress,4.6) +- (0,0.0) 
      };

      \addplot+[ybar,error bars/.cd,y dir=both,y explicit] plot 
      coordinates {
        (gnugo-dfl,8.6) +- (0,0.04)
        (gnugo-dfl-stress,11.7) +- (0,.1) 
        (gnugo-our,6.8) +- (0,0.03)
        (gnugo-our-stress,8.7) +- (0,0.1) 
        (stream-dfl,77.3) +- (0,0.02)
        (stream-dfl-stress,79.7) +- (0,.1) 
        (stream-our,72.2) +- (0,0)
        (stream-our-stress,76.2) +- (0,0.6) 
        (syscalls-dfl,63.5) +- (0,0.4)
        (syscalls-dfl-stress,62.4) +- (0,0.1) 
        (syscalls-our,47.5) +- (0,0.1)
        (syscalls-our-stress,48.2) +- (0,0.1) 
      }; 

      \addplot+[ybar,error bars/.cd,y dir=both,y explicit] plot 
      coordinates {
        (gnugo-dfl,34.4) +- (0,0)
        (gnugo-dfl-stress,32.1) +- (0,0.1) 
        (gnugo-our,35.1) +- (0,0.01)
        (gnugo-our-stress,33.4) +- (0,0.2) 
        (stream-dfl,17.4) +- (0,0)
        (stream-dfl-stress,14.8) +- (0,.4) 
        (stream-our,21.9) +- (0,0)
        (stream-our-stress,18.4) +- (0,0.3) 
        (syscalls-dfl,14.5) +- (0,0.2)
        (syscalls-dfl-stress,14.8) +- (0,0.1) 
        (syscalls-our,26.3) +- (0,0)
        (syscalls-our-stress,26.1) +- (0,0.0) 
      }; 

      \legend{\strut FE, \strut BAD, \strut BE, \strut RET}
    \end{axis}
  \end{tikzpicture}